\documentclass{JHEP3}
\usepackage{epsfig}
\usepackage{graphicx}
\usepackage{amssymb}
\usepackage{subfigure}
\newcommand{\beq}{\begin{eqnarray}}
\newcommand{\eeq}{\end{eqnarray}}

\title{Moduli Stabilization and Supersymmetry Breaking in Deflected Mirage Mediation}

\author{Lisa~L.~Everett, Ian-Woo Kim, Peter Ouyang, and Kathryn M.~Zurek \\
Department of Physics, University of Wisconsin,
Madison, WI 53706, USA}

\preprint{MADPH-08-1513\\ MADTH-08-08}

\abstract{We present a model of supersymmetry breaking in which the contributions from gravity/modulus, anomaly, and gauge mediation are all comparable. We term this scenario ``deflected mirage mediation," which is a generalization of the KKLT-motivated mirage mediation scenario to include gauge mediated contributions.  These contributions deflect the gaugino mass unification scale and alter the pattern of soft parameters at low energies.  In some cases, this results in a gluino LSP and light stops; in other regions of parameter space, the LSP can be a well-tempered neutralino.  We demonstrate explicitly that competitive gauge-mediated terms can naturally appear within phenomenological models based on the KKLT setup by addressing the stabilization of the gauge singlet field which is responsible for the masses of the messenger fields.   For viable stabilization mechanisms, the relation between the gauge and anomaly contributions is identical in most cases to that of deflected anomaly mediation, despite the presence of the K\"{a}hler modulus.  Turning to TeV scale phenomenology, we analyze the renormalization group evolution of the supersymmetry breaking terms and the resulting low energy mass spectra.   The approach sets the stage for studies of such mixed scenarios of supersymmetry breaking at the LHC.}

\begin{document}

\section{Introduction and Motivation}

Low energy (electroweak to TeV scale) softly broken supersymmetry (SUSY) (for recent reviews, see \cite{Martin:1997ns,Chung:2003fi}) has long been considered to be the best-motivated candidate for physics beyond the Standard Model (SM), due to its elegant resolution of the hierarchy problem, radiative mechanism for electroweak symmetry breaking, and predicted dark matter candidate (assuming a conserved R-parity).  Theories with low energy supersymmetry, such as the minimal supersymmetric standard model (MSSM),  will shortly face unprecedented experimental tests at the Large Hadron Collider (LHC) at CERN.  As the phenomenology of such theories is dictated by the superpartner mass spectrum, which in turn is governed by the soft supersymmetry breaking sector, understanding how SUSY is broken is the most important question for studies of low energy supersymmetry at the LHC.

In viable models of supersymmetry breaking, supersymmetry is broken in a hidden or secluded sector and is transmitted to the observable sector by mediator fields, which develop auxiliary component (F and/or  D term) vacuum expectation values (vevs).  The mediators generically couple to SM fields via loop-suppressed and/or nonrenormalizable interactions.  In each model, the characteristic spectrum of soft masses is set by: (i) the specific mediation mechanism of SUSY breaking,  (ii) which mediators get the largest auxiliary field vevs, and (iii) the dominant effects which produce the couplings between the mediators and the SM fields.  Hence, as supersymmetry breaking itself typically occurs at a high scale, the low scale pattern of soft masses may tell us something about the high scale physics, not  directly reachable by experiments, which determines these parameters.

Phenomenological studies of low energy supersymmetry largely focus on models in which one mediation mechanism is dominant (see e.g.~\cite{Allanach:2002nj});  in the bottom-up approach, this is often to solve a given phenomenological problem of the MSSM (such as the $\mu$ problem, the flavor/CP problems, etc.). Such models can be roughly classified into gravity mediated, gauge mediated, and  (braneworld-motivated) ``bulk" mediated models.  Gravity mediated terms\cite{grav1}, which arise from couplings that vanish as the Planck mass $M_P\rightarrow \infty$,  include modulus mediation contributions  \cite{modulus} and (loop-suppressed) anomaly mediation terms \cite{anomaly}, among others.  Gauge mediated terms arise from loop diagrams involving new messenger fields with SM charges \cite{gauge1,gauge2,Giudice:1998bp}, whereas bulk mediated terms arise from bulk mediator fields in braneworld scenarios, such as gaugino mediation \cite {gaugino} and $Z'$ mediation \cite{zprime}.  Since gravity is a bulk field, certain gravity mediated models, which include the pure anomaly mediation scenario (which requires sequestering), are also bulk mediation models.

A complementary approach is to consider models in which more than one mediation mechanism plays an important role.  Such scenarios can be motivated within the top-down, string-motivated approach to supersymmetry breaking; the prototype example of this type is mirage  mediation~\cite{Choi:2004sx,Choi:2005ge}, which is motivated from the Kachru-Kallosh-Linde-Trivedi (KKLT) approach to moduli stabilization within Type IIB string theory  \cite{Kachru:2003aw}.  In mirage mediation, the modulus mediated supersymmetry breaking terms are suppressed by $\ln M_{pl}/m_{3/2}$, which is numerically of the order of a loop factor, such that the anomaly mediated terms are competitive.  This results in mirage unification, in which the gaugino and scalar masses unify at a scale much below where the soft masses are generated.  Mirage mediation has distinctive phenomenological features \cite{Choi:2005uz,miragepheno,Choi:2006bh,Choi:2008hn}, including a non-standard gaugino mass  pattern \cite{Choi:2007ka}  and reduced low energy fine-tuning \cite{littlehierarchy}.

In this spirit, we recently outlined a framework in which the KKLT-motivated mirage mediation scenario was extended to incorporate three of the most-studied supersymmetry breaking mechanisms: modulus mediation, anomaly mediation, and gauge mediation \cite{Everett:2008qy} (for a similar scenario, see \cite{Nakamura:2008ey}).  We argued in \cite{Everett:2008qy} that  the necessary ingredients for gauge mediated contributions can naturally be present within the KKLT setup  and demonstrated the gauge mediated terms are indeed comparable to the modulus and anomaly mediated terms, leading to a scenario we denoted as {\it deflected mirage mediation}.  The gauge-mediated terms deflect the soft terms from their mirage mediation renormalization group trajectories, in analogy with deflected anomaly mediation \cite{Pomarol:1999ie,Rattazzi:1999qg} (for a more recent analysis see \cite{Endo:2008gi}).  This effect deflects the gaugino mass unification scale and modifies the values of the soft terms at low energies.    As in mirage mediation, the pattern of soft masses depends on the ratios of the contributions from the different mediation mechanisms.  These ratios, which are parameters of the model, generically take on discrete values in string-motivated scenarios in which the stabilization of the mediator fields is addressed.  

This paper is a companion to our previous paper \cite{Everett:2008qy}, in that here we provide a detailed discussion of deflected mirage mediation.  Our primary goal is to demonstrate explicitly that the approach is on firm theoretical ground within the KKLT-motivated setup.  In particular, we will show that the mediator of gauge mediation can acquire an F term vev purely through supergravity effects,  eliminating the need for a separate dynamical supersymmetry breaking sector.  This leads to competitive contributions to the observable sector soft terms from gauge mediation to the previously known mirage mediation results.   Since this mediator field is a matter modulus which requires stabilization, we consider several possible stabilization mechanisms, including stabilization through radiative supersymmetry breaking effects and stabilization by higher-order superpotential self-couplings, which have previously been considered in the context of deflected anomaly mediation \cite{Pomarol:1999ie}.  We find the interesting result that despite the presence of the K\"{a}hler modulus, the ratio between the gauge mediated and anomaly mediated contributions is identical in most cases to what was found in deflected anomaly mediation.   With this in hand, we compute the MSSM soft terms, and study the renormalization group trajectories of the parameters and the resulting low energy mass patterns.   The analysis sets the stage for further phenomenological studies of low energy supersymmetry within this general framework.

The outline of this paper is as follows.  In Section~\ref{theorysect}, we discuss the theoretical motivation and model-building aspects of deflected mirage mediation, focusing on the stabilization of the mediator fields. We compute the soft supersymmetry breaking parameters of the MSSM fields in Section~\ref{susybrksect}.
The phenomenological implications of this string-motivated scenario are presented in Section~\ref{phenosect}.  Our analysis includes an investigation of the renormalization group running of the parameters and sample spectra for theory-motivated choices of the parameters, including the phenomenologically interesting case in which all contributions are  roughly of the same size.   In Section~\ref{concl}, we provide our conclusions and outlook.

%%%%%%%%%%%%%%%%%%%%%%%%%%%
\section{Moduli Stabilization and Supersymmetry Breaking}
\label{theorysect}

\begin{figure}[t]
\begin{center}
{\epsfig{figure=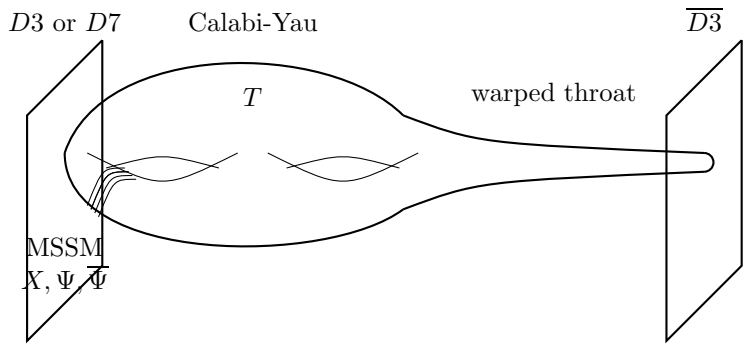}
\caption{The KKLT-motivated braneworld setup for deflected mirage mediation. \label{KKLTfig}}}
\end{center}
\end{figure}

In this section, we provide a detailed discussion of the theoretical background and model-building aspects of deflected mirage mediation within the context of the KKLT approach to moduli stabilization, which has been a primary motivation for mirage mediation models.    Recall that in the KKLT construction \cite{Kachru:2003aw}, which was inspired by various results in Type IIB string theory flux compactifications (see e.g.~\cite{Giddings:2001yu}), four-dimensional $\mathcal{N}=1$ supersymmetry is broken by anti-branes located at the tip of the warped throat geometry produced by three-form fluxes, while the observable sector arises from stacks of D branes located in the bulk Calabi-Yau space, as shown schematically in Figure~\ref{KKLTfig}.  

In phenomenological models based on this construction, it was found \cite{Choi:2004sx,Choi:2005ge} that the observable sector soft SUSY breaking terms are dominated by two mediation mechanisms: (i) modulus mediation, due to the K\"{a}hler modulus $T$, and (ii) anomaly mediation, which can be represented by the conformal compensator field $C$ of the gravity multiplet.     As the scale of observable sector soft terms due to modulus mediation is of the order 
\begin{equation}
\label{graveq}
m_{\rm soft}^{({\rm modulus})} \sim  \frac{F^T}{T+\overline{T}},
\end{equation}
while that of the (loop-suppressed) anomaly mediated terms is of the order
\begin{equation}
m_{\rm soft}^{({\rm anomaly})}\sim \frac{1}{16\pi^2} \frac{F^C}{C},
\end{equation}
the comparable size of the two scales indicates a suppression of the modulus F term; the suppression factor was determined to be $\ln(M_P/m_{3/2}) \sim 4\pi^2$  \cite{Choi:2004sx,Choi:2005ge}.

In deflected mirage mediation, this KKLT-motivated picture is generalized to include additional observable sector superfields, a gauge singlet $X$ and  $N$ vectorlike pairs of ``messenger" fields $\Psi$, $\overline{\Psi}$ with SM gauge charges, as shown in Figure~\ref{KKLTfig}.   Such states are generically present in string theory models along with the MSSM fields, so their inclusion is not without top-down motivation. In this string-motivated context, $X$ represents an open string mode which starts and ends on observable sector branes, not an open string mode which connects observable and hidden sector branes.  $X$ is thus sequestered from the hidden sector in the same way as the MSSM matter content.  More generally, the $X$ field can thus be regarded as a matter modulus which must be also stabilized by some mechanism.  

Interestingly, if the stabilization mechanism is mainly due to supersymmetry breaking effects, then the supersymmetry breaking order parameter $F^X/X$ can have the same order of magnitude as other SUSY breaking order parameters.   We will demonstrate that $X$ can acquire an F term vacuum expectation value purely due to supergravity effects, such that the separate dynamical supersymmetry breaking sector typically needed in gauge mediation models is not necessary. This leads to an additional contribution to the observable sector soft terms through gauge-mediated messenger loops: 
\begin{equation}
m_{\rm soft}^{({\rm gauge})}\sim \frac{1}{16\pi^2} \frac{F^X}{X}.
\end{equation}
We will also show that  in a broad class of models, 
\begin{equation}
\frac{F^X}{X}\sim \frac{F^C}{C},
\end{equation}
and hence $m_{\rm soft}^{({\rm gauge})}$ can generically be comparable to $m_{\rm soft}^{({\rm anom})}$ and $m_{\rm soft}^{({\rm grav})}$.  Our goal in this paper is simply to demonstrate that all three mediation mechanisms
can in principle coexist, although more detailed model-building would certainly be interesting.  To this end, in what follows we will address the stabilization of $X$ and $T$ within the context of the four-dimensional effective supergravity theory of this KKLT-motivated scenario.\footnote{Mostly, we will use units in which the reduced Planck mass $M_P= 2.4 \times 10^{18} \mbox{ GeV}$ is set to unity, but  will restore it occasionally when discussing issues regarding orders of magnitude.}  \\

\subsection{Theoretical Background}
We begin with the four-dimensional $\mathcal{N}=1$ effective supergravity within this class of KKLT constructions, which can be expressed concisely in the chiral compensator formalism,
\begin{equation}
{\cal L} = 
\int d^4 \theta C \overline{C} G( \Phi, \overline{\Phi} ) +\left[ \int d^2 \theta \frac{1}{4} f_a (\Phi) W^{a\,\alpha} W^a_\alpha 
+ \int d^2 \theta C^3 W(\Phi) +\mbox{h.c.}
 \right],
 \label{ccformalism}
 \end{equation}
in which $C$ is the chiral compensator and $\Phi$ denotes a generic chiral superfield.  In Eq.~(\ref{ccformalism}),
\begin{eqnarray}
G( \Phi, \overline{\Phi} ) = -3 \exp ( -K(\Phi, \overline{\Phi}) / 3 ),
\end{eqnarray}
where $K$ is the K\"{a}hler potential, $W^a_\alpha$ are the (generically $\Phi$-dependent)  gauge field strength superfields of the SM gauge groups $G_a = \{SU(3)_C,\,SU(2)_L,\,U(1)_Y\}$, and  $W(\Phi)$ is the superpotential.   The vev of the gauge kinetic function $f_a$ (which can be field-dependent) is related to the gauge coupling $g_a$ and $\theta_a$ parameter via
\begin{eqnarray}
\langle f_a \rangle = \frac{1}{g^2_a} + i \frac{\theta_a}{8 \pi^2}.
\end{eqnarray}
As stated previously, we study the scenario in which $\Phi$ includes the K\"{a}hler modulus $T$ associated with the volume of the compact space, the gauge singlet $X$, the $N$ messenger pairs $\Psi$, $\overline{\Psi}$ (usually taken to be complete multiplets under a grand unified group which contains the SM gauge group to preserve gauge coupling unification), and the MSSM fields, which we now denote as $\Phi_i$.   Furthermore, we consider the situation in which only $T$, $C$, and $X$ develop sizable F term vacuum expectation values.

The  K\"{a}hler potential can be expanded in powers of $\Phi_i$ according to
\begin{equation}
\label{fullkahler}
K = K_0 ( T )  + Z_{X} (T, \overline T ) X \overline X + Z_{i} (T,\overline{T}) \Phi_i \overline{\Phi}_i + 
{\mathcal O} \left((|\Phi|^4, |X|^4)\right), 
\end{equation}
%ADD Y definition later.
in which $K_0$ is the K\"ahler potential of $T$, and $Z_X$ ($Z_i$) is the K\"ahler metric 
of $X$ ($\Phi^i$).  In Eq.~(\ref{fullkahler}), we have assumed a diagonal matter metric for simplicity.  Higher order terms are ignored, since the vevs of $X$ and $\Phi^i$ are assumed to be negligible compared with the Planck scale.   At leading order, $K_0$  generically is assumed to have the ``no-scale" form
\begin{equation}
\label{noscaleK}
K_0( T, \overline{T} ) = - 3 \log (T+\overline{T}).
\end{equation}
To see deviations from this form, we will use the following generalization:
\begin{equation}
\label{generalK0}
K_0( T, \overline{T} ) = - p \log (T+\overline{T}),
\end{equation}
such that we can examine the results with the no-scale form by setting $p=3$.
The K\"{a}hler metrics $Z_X(T,\overline{T})$ and $Z_i(T,\overline{T})$ are dictated by the geometric scaling behavior of $X$ and $\Phi_i$ with respect to changes in the overall compactification volume. $Z_X$ and $Z_i$ are given by
\begin{equation}
\label{kahlermet}
Z_X=\frac{1}{(T+\overline{T})^{n_X}},\;\;\;Z_i=\frac{1}{(T+\overline{T})^{n_i}},
\end{equation}
in which $n_X$ and $n_i$ are the modular weights of $X$ and $\Phi_i$, respectively. The superpotential takes the form
\begin{equation}
\label{superpot}
W=W_0+W_1(X)+\lambda X\Psi\overline \Psi +W_{\rm MSSM}, 
\end{equation}
in which $W_0$ is the stabilizing superpotential for $T$, $W_1(X)$ denotes the singlet self-interaction superpotential terms, and $W_{\rm MSSM}$ is the standard MSSM superpotential.  Here we have assumed a renormalizable superpotential coupling among $X$ and the messenger pairs; a generalization to higher order interaction terms is straightforward.  $W_0$ is given by
\begin{eqnarray}
W_0 = w_0 - A e^{-a T},
\label{W0}
\end{eqnarray}
in which $w_0$ is a constant superpotential term determined by the flux compactification, and the second term is a nonperturbative contribution which arises from gaugino condensation or $D3$-instanton corrections, such that $a\sim {\mathcal O} (8\pi^2)$ and $A\sim {\mathcal O} (M_{\rm P}^3)$.   In the standard KKLT model, one makes the further assumption that the perturbatively generated $w_0$ can be chosen to
be small enough that the nonperturbative term is competitive with it for moderately large $T$. This assumption justifies the neglect of higher string corrections, and allows us to use the no-scale K\"{a}hler potential of Eq.~(\ref{noscaleK}).
We will study three possibilities for the functional form of the singlet self-interaction $W_1(X)$: 
\begin{itemize}
\item The singlet $X$ may have a mass term, such that at leading order
\begin{equation}
\label{susyxint}
W_1(X)=\frac{1}{2}m (X-X_0)^2,
\end{equation}
where $X_0$ is the vacuum expectation value of $X$.  This corresponds to supersymmetric stabilization via F terms; one can also consider D term stabilization, such as through an anomalous $U(1)$, as discussed in  \cite{Choi:2006bh}.
%in which the $\ldots$ denote higher order terms.
\item  The singlet may have no self-interactions, such that 
\begin{equation}
\label{noxint}
W_1(X)=0.
\end{equation}
In this case, $X$ is stabilized due to radiative corrections to the $X$-dependent supersymmetry breaking potential.
\item The self-interactions may appear at renormalizable or higher order, or can be generated by a nonperturbative mechanism:
\begin{equation}
\label{higherxint}
 W_1(X) = \frac{X^n}{\Lambda^{n-3}},
 \end{equation}
 in which $\Lambda$ is the scale at which these terms are generated.  For perturbatively generated terms, $\Lambda$ is the cutoff scale, while for nonperturbatively generated terms, $\Lambda$ is the scale associated with dynamical symmetry breaking.
 \end{itemize}
  
 In the KKLT setup, the combination of the two terms of the stabilizing potential $W_0$  for $T$ leads to a supersymmetric minimum with a negative cosmological constant. To break supersymmetry and cancel the cosmological constant, anti-D3 branes are put at the tip of the warped throat.  Such breaking effects are encoded in Lagrangian terms with nonlinearly realized supersymmetry.  The set of such terms relevant for low energy supersymmetry breaking is given by the following contribution to the scalar potential:
 \begin{equation}
 {\mathcal L}_{\rm (NL)} \approx -\int d^4 \theta (C \overline{C})^2
{\mathcal P} \theta^2 \bar{\theta}^2,
\label{nonlinear2}
 \end{equation}
in which $P$ generically can be a function of the superfields $\Phi$.  In the KKLT construction, $P$ is effectively constant due to the warped geometry.  Here we will generalize this form to 
\begin{equation}
{\mathcal P} \sim (T+\overline{T})^{n_P}.
\label{np}
\end{equation}
The nonlinearly realized sector is assumed to be well sequestered from the observable sector matter fields (e.g.~$X$, $\Phi_i$).  Eq.~(\ref{np}) leads to a so-called uplifting potential of the form
\begin{equation}
V_L= \frac{ \mathcal{D} }{(T+\bar{T})^{2-n_P}}.
\label{uplift}
\end{equation}
In the KKLT model, $n_P=0$ and $\mathcal{D}$ is the warped string scale at the tip
of the throat. 

\subsection{Review of KKLT Modulus Stabilization}
Let us briefly review the stabilization of $T$, assuming that only $T$ plays an important role (further details can be found in the mirage mediation literature \cite{Choi:2004sx,Choi:2005ge}).   Heuristically speaking, the argument is  as follows.  In the absence of the uplifting potential of  Eq.~(\ref{uplift}), $T$ is stabilized at a supersymmetric minimum due to the interplay of the flux-generated and nonperturbative terms in  $W_0$ ($T$ would be a flat direction in the absence of the nonperturbative term; a runaway behavior would result if $w_0=0$).   The modulus vev $t_0$ thus satisfies the F-flatness condition:
\begin{equation}
\label{flatness}
D_T W_0\equiv \partial_T W _0+ K_T W_0=0,
\end{equation}
in which $K_T=\partial_T K$ and $W_0\neq 0$.  
This results in a supersymmetric AdS vacuum, with vacuum energy given by $-3m_{3/2}^2M_P^2$.  As we will see explicitly below, the uplifting potential shifts the vacuum expectation value of $T$ by an amount of order
\begin{equation}
\label{vacshift}
\frac{\Delta T}{t_0}\sim \mathcal{O}\left( \frac{m^2_{3/2}}{m^2_T} \right),
\end{equation}
in which the mass of $T$
\begin{equation}
\label{tmod}
m_T\simeq - \left . \frac{e^{K/2}\partial_T^2W}{\partial_T\partial_{\overline{T}}K} \right |_{T=t_0}
\end{equation}
will turn out to be parametrically larger than the gravitino mass $m_{3/2}=e^{K/2}|W|$.  This vacuum shift induces an F term for $T$ of the order
\begin{equation}
\frac{F^T}{T+\overline{T}}\sim \mathcal{O}\left (\frac{m^2_{3/2}}{m_T} \right ).
\end{equation}
Since $F^T/(T+\overline{T})$ dictates the size of the observable sector supersymmetry breaking, the mass of the $T$ modulus governs the extent to which modulus mediation contributes to the soft terms.  Hence, the F terms of moduli with masses of order the string scale, such as the dilaton and complex structure moduli, are irrelevant for TeV scale phenomenology.  

To determine $F^T/(T+\overline{T})$ and $m_T$, we take the standard approach of treating the uplifting potential as a perturbation about the supersymmetric minimum; this procedure is valid since $T$ generically has a large vacuum expectation value, and the curvature of the potential at this minimum is governed by $m_T\gg m_{3/2}$. The unperturbed minimum is determined from Eq.~(\ref{flatness}) (using Eqs.~(\ref{generalK0}) and (\ref{W0}) and assuming a real $t_0$) as follows:
\begin{equation}
D_TW_0=\left. a A e^{-a T} - \frac{p}{(T+\bar{T})} (w_0 - A e^{-a T})
\right|_{T=t_0 }
\end{equation}
The solution for $t_0$ is given by the expression
\begin{equation}
a t_0 = - {\mathcal W} \left( -\frac{p w_0}{2 A e^{p/2}} \right)
- \frac{p}{2}, \label{vevofT}
\end{equation}
in which ${\mathcal W}(x)$ is the Lambert ${\mathcal W}$ function, which is the solution $z$ of the equation $x = z e^z$. 
%Roughly, if the argument $x$ of ${\mathcal W}(x)$ is either exponentially small or large, ${\mathcal W}(x)$ is
%similar to the logarithm of $x$. 
Note that if $w_0$ is exponentially small,  Eq.~(\ref{vevofT}) reduces to 
\begin{equation}
\label{approxat0}
a t_0 \simeq  \ln{ \left ( \frac{A}{w_0}\right )}. 
\end{equation}
The gravitino mass is 
\begin{eqnarray}
m_{3/2} &=& e^{K_0 /2} W_0 = \frac{1}{ (2 t_0)^{p/2}}
\left( w_0 - A e ^{-a t_0 } \right)
% \nonumber \\
%&\approx&
\approx  \frac{1}{(2 t_0 )^{p/2}} w_0,
\label{gravitino}
\end{eqnarray}
in which the last approximate equality holds if $a t_0 \gg p/2$.  Since this expression should not change in the presence of the uplifting potential, Eq.~(\ref{approxat0}) yields
\begin{equation}
a t_0 \sim \log ( M_P / m_{3/2} ),
\end{equation}
recalling that $A\sim M_P^3$ and $w_0\sim m_{3/2}M_P^2$.  From Eq.~(\ref{tmod}) (still neglecting the uplifting potential), the mass $m_T$ is of the order
\begin{equation}
m_T \approx  2 a t_0  m_{3/2},
\end{equation}
which demonstrates that $m_T$ is indeed parametrically heavier than the gravitino mass.

Turning now to the effects of the uplifting potential, one must keep in mind that the vacuum expectation value of $V_L$ must be of the order $3m^2_{3/2}M_P^2$ to cancel the negative cosmological constant of the supersymmetric AdS minimum.   Expanding $T$ about the minimum $t_0$ given in Eq. (\ref{vevofT}), the scalar potential takes the form
\begin{eqnarray}
\left. V  \right|_{t_0 + \Delta t } = 
\frac{p}{4t_0^2} m_t^2 \left(\Delta t\right)^2
-3 (2-n_p) m_{3/2}^2 \left( \frac{\Delta t}{t_0} \right).
\end{eqnarray} 
Minimizing this potential with respect to $\delta t$ leads to 
\begin{equation}
\left. \frac{\partial V}{\partial \Delta t} \right|_{t_0 + \delta t} 
=\frac{p m_T^2 }{2 t_0^2} \Delta t 
- \frac{3 (2 - n_P)}{t_0} m_{3/2}^2 = 0, 
\end{equation}
such that
\begin{equation}
\frac{\Delta t}{t_0} = \frac{6(2-n_P)}{p} \frac{m_{3/2}^2}{m_T^2}.
\end{equation}
The shift in the vacuum expectation value of $T$ induces an F term of the form
\begin{eqnarray}
F^T = -e^{K_0 /2} (K_0)^{T\overline{T}} D_{\overline{T}} W^* = \frac{6(2-n_P)}{p} \frac{m^2_{3/2}}{m_T} t_0. 
\end{eqnarray}
The compensator also gets a nonzero F term
when supersymmetry is broken,
\begin{equation}
F^C = \frac{\bar{C}^2}{3}a A e^{-aT} \simeq C\frac{w_0}{(T+\overline{T})^{p/2}}.
\label{compensatorf}
\end{equation}
Comparing Eq.~(\ref{compensatorf}) with Eq.~(\ref{gravitino}) demonstrates that the gravitino mass is approximately given by $m_{3/2} \approx F^C/C$.  Hence,
\begin{eqnarray}
\frac{F^T}{T+\overline{T}} = \frac{3(2-n_p)}{2p} \frac{1}{a t_0} \frac{F^C}{C}.
\end{eqnarray}
In the KKLT case with $p=3$ and $n_p=0$, 
\begin{eqnarray}
\frac{F^T}{T+\overline{T}} = \frac{1}{a t_0} \frac{F^C}{C}.
\label{modulusanomaly}
\end{eqnarray}
Eq.~(\ref{modulusanomaly}) shows that the modulus mediation contribution $F^T/(T+\bar{T})$ is suppressed with respect to $F^C/C$ by $a t_0 \simeq \ln(M_P/m_{3/2})  \sim 4 \pi^2$.  Since this factor is numerically of the order of a loop suppression, the modulus and anomaly mediated soft terms are comparable, which is the standard mirage mediation result.\\

\subsection{Stabililization of the Matter Modulus $X$}
To generalize to the case in which $X$ and the messenger pairs are present, we will now address the issue of the stabilization of $X$ within each of the three previously discussed stabilization mechanisms.\\

%\subsubsection*{Supersymmetric Stabilization at a High Scale.}
\noindent $\bullet$ {\bf Supersymmetric Stabilization at a High Scale.} 
We first consider the case in which $X$ is stabilized by a supersymmetric mechanism at a high scale.  One example is F term stabilization through an explicit superpotential mass term of the form
$W=\frac{1}{2}m_X(X-X_0)^2$ as in Eq.~(\ref{susyxint}), which may be thought of as the leading term in a Taylor expansion of the potential.   For this term to play an important role in the stabilization of $X$, the mass parameter $m_X$ must be hierarchically larger than the gravitino mass, and the vacuum expectation value $X_0$ should be less than the Planck mass.  In the absence of the uplifting potential $V_L$, the F term of $X$ clearly vanishes.  If  supersymmetry breaking effects acted to shift the vev of $X$,  a nonzero $F^X$ would be induced (as in the case of $T$).  However, here such effects do not lead to a shift in $X$.  For example, the soft supersymmetry breaking $B$ term associated with  Eq.~(\ref{susyxint})  generated from anomaly mediation has the form ${\mathcal O}(m_{3/2} m_X (X-X_0)^2)$, which keeps the vev of $X$ at $X_0$; higher-order terms will also maintain $X=X_0$ as a (meta-)stable vacuum.  Therefore, if $m_X\gg m_{3/2}$, $F^X$ will be zero, resulting in no gauge-mediated contributions to the MSSM soft terms.

The situation is more involved when considering D term stabilization via Fayet-Iliopoulos (FI) terms, which can result in supergravity either from (i) gauged $U(1)_R$ symmetries or (ii) anomalous $U(1)$ gauge groups.  The case of the anomalous $U(1)$ was studied in \cite{Choi:2006bh}, we summarize their results here for completeness. Such anomalous $U(1)$ gauge groups are ubiquitous in string models.  Their anomalies are canceled at one-loop by the Green-Schwarz mechanism, which manifests itself in the low energy effective field theory as a nonlinear realization of the $U(1)$ symmetry, under which a subset of the moduli (the Green-Schwarz moduli) transform.  This effect breaks the $U(1)$ gauge symmetry at a scale given by the string scale divided by a loop  factor, and leads to a moduli-dependent FI term.  Hence, if $X$ is charged under the anomalous $U(1)$, it can be triggered to acquire a vev by the FI term.  In this case, the $X$ degree of freedom is absorbed in the vector multiplet of the anomalous $U(1)$, and $F^X$ can be induced if the Green-Schwarz modulus has an F term.  In KKLT models, $T$ can be a Green-Schwarz modulus, but the induced $F^X/X$ is ${\mathcal O} (F^T / (T+\overline{T}))$, which gives a subleading contribution to the soft terms  \cite{Choi:2006bh}.

In summary, stabilizing $X$ at a high scale either by F terms or D terms does not yield a situation in which $F^X/X$ results in comparable soft terms to that of mirage mediation.  Hence, we now turn to the other two stabilization mechanisms.\\

%\subsubsection*{Radiative Stabilization.}
\noindent $\bullet$ {\bf Radiative Stabilization.}
We now consider the case in which $X$ is massless before supersymmetry breaking.   If $X$ has no superpotential self-interactions, as in Eq.~(\ref{noxint}),  a purely nonsupersymmetric stabilization results because once supersymmetry is broken, $X$ should acquire a soft mass from anomaly mediation, lifting its flat direction. If this soft mass term for $X$ changes sign at a particular scale due to renormalization group running,  $X$ is stabilized at a value of the order of this crossover point.  More precisely, since $X$ couples to the vectorlike matter pairs $\Psi$, $\overline{\Psi}$ with coupling strength $\lambda$ (see Eq.~(\ref{superpot})), 
\begin{equation}
V(X) \sim \frac{1}{16\pi^2} \left| \frac{F^C}{C} \right|^2 
N \lambda(X)^2 \left[ A_0 \lambda^2(X) - C_a g^2_a(X) \right] |X|^2,
\end{equation}  
for some positive $A_0$ and $C_a$.  One can think of this potential as
being generated by the Coleman-Weinberg mechanism \cite{Coleman:1973jx}; it arises from integrating out the messengers
$\Psi$ at one loop.  Depending on $\lambda$ and $g_a$, $X$ can be stabilized
at a scale anywhere between the electroweak scale and the UV cutoff scale \cite{Pomarol:1999ie}.  The vev that sets the mass also sets the scale of the messengers $\Psi$, as seen in Eq.~(\ref{superpot}).

The F term of $X$ induced by this radiative stabilization mechanism can be simply obtained by
\begin{eqnarray}
F^X \simeq - e^{K_0/2} K^{X\bar{X}} D_{\bar{X}} W\simeq - e^{K_0/2} K^{X\bar{X}} K_{\bar{X}} W_0  \simeq  -m_{3/2} X,
\end{eqnarray}
such that 
\beq
\frac{F^X}{X} = - \frac{F^C}{C} \approx -m_{3/2}.
\label{radstab}
\eeq
The contribution to the observable sector soft terms from gauge-mediated interactions involving messenger loops is thus comparable to the anomaly-mediated terms.  This result is identical to that of the deflected anomaly mediation scenario of Pomarol and Rattazzi \cite{Pomarol:1999ie}, despite the nontrivial K\"{a}hler potential of $X$ and $T$.  The feature that the F terms for this model are equal in magnitude but opposite in sign is distinctive; as a result, deflected anomaly mediation is also called ``anti-gauge mediation" \cite{Pomarol:1999ie}.  As we will see shortly, this relation can be modified by giving $X$ a more complicated superpotential. 
It is also worthwhile to note that the standard mirage mediation F term constraints for $T$ and $C$ are only modified by Planck-suppressed corrections, provided that $F^{\bar{X}}$ and $F^{\bar{C}}$ are
comparable, and also that $X$ is stabilized below the Planck scale.\\

%\subsubsection*{Higher-order Stabilization.}
\noindent $\bullet$ {\bf Higher-order Stabilization.}
Let us finally consider the case in which $X$ is stabilized by the interplay between 
the scalar potential induced by supersymmetry breaking effects and higher-order $X$ self-interaction superpotential terms, as represented by $W_1$ in Eq.~(\ref{superpot}).  This situation was also considered in \cite{Pomarol:1999ie} for the case of deflected anomaly mediation.  Here we limit ourselves to the case in which only a single term in $W_1$ of the form of  Eq.~(\ref{higherxint})  dominates the stabilization:
\begin{equation}
W_1 =  \frac{ X^n }{\Lambda^{n-3}} + \dots, 
\end{equation}
in which $\Lambda$ is the cutoff scale.  The exponent $n$ can have an arbitrary value which may be positive or negative; a negative exponent would indicate that this term originates from nonperturbative dynamics. 
The  K\"ahler potential terms which are relevant for the stabilization of $X$ are the leading order contributions for $T$ and $X$ as given in Eq.~(\ref{fullkahler}), Eq.~(\ref{generalK0}), and Eq.~(\ref{kahlermet}).
Revealing the comformal compensator dependence, the K\"ahler terms of the superspace action are
\begin{eqnarray}
\int d^4 \theta  G &=&% \int d^4 \theta C C^* \Omega 
 \int d^4 \theta \left( -3 C \overline{C} e^{-K/3} \right)  \nonumber\\ &\simeq&
-3 C\overline{C} (T+\overline{T})^{p/3} + C \overline{C} \frac{1}{(T+\overline{T})^{n_X-p/3}}
X \overline{X} 
\nonumber\\
&\simeq&
 -3 C \overline{C} (T+\overline{T})^{p/3} + \frac{1}{(T+\overline{T})^{n_X - p/3}}
\tilde{X} \overline{\tilde{X}},
\end{eqnarray}
in which $X$ is holomorphically redefined to $\tilde{X}  = C X$ to simplify the K\"ahler potential.\footnote{ Our definition of $X$ and $\tilde{X}$ is opposite to that of \cite{Pomarol:1999ie}.}
%simpler without mixing between $C$ and $\tilde X$.
The scalar potential is of the form
\begin{eqnarray}
V = G_{A\overline{B}} F^A \overline{F}^{B},
\end{eqnarray}
in which the indices $A,B$ run over $C,\,T,\,\tilde{X}$, and $G_{A\overline{B}} \equiv\partial_A\partial_{\bar{B}}G$.  The F terms are given by 
\begin{eqnarray}
F^A = -G^{A\overline{B}} \partial_{\overline{B}}(C^3 W),
\end{eqnarray}
in which $G^{A\overline{B}}$ is the inverse of $G_{A\overline{B}}$.  We pause to comment here that the F term equations for the messengers impose the condition that the messenger vevs must vanish; these constraints can be consistently satisfied after supersymmetry breaking as well. Hence, the messenger fields and their F-terms can be safely set to zero in the following analysis.

In the calculable scenario in which the vev of $X$ is smaller than $M_P$, such that higher-order K\"{a}hler potential terms for $X$ can be safely neglected, $X$ and $F^X$ can be treated as small numbers, and $T$, $F^T$ and $F^C$ can be perturbed about the values which were obtained in the previous subsection:
\begin{eqnarray}
F^C &=& F_0^C + \delta F^C, \nonumber \\
F^T &=& F_0^T + \delta F^T,
\end{eqnarray}   
in which $F_0^C$ and $F_0^T$ are the F terms of $C$ and $T$ in the absence of $X$: 
\begin{eqnarray}
F^C_0 &\approx& \overline{C}^2 e^{K_0/3} \overline{W}_0 = \frac{\overline{C}^2 \overline{W}_0}{ (T+\overline{T})^{p/3}}, \nonumber\\
\frac{F^T_0}{T+\overline{T}} &=& \frac{2}{a(T+\overline{T})} \frac{F^C_0}{C}.
\label{zerorelation} 
\end{eqnarray}
Therefore,  we obtain
\begin{eqnarray}
\delta F^C &=& \frac{1}{9} (p-3)(n-3) 
\frac{\overline{\tilde{X}}^n}{\Lambda^{n-3} \overline{C}^{n-2} (T+\overline{T})^{p/3}}
-\frac{p-3n_X}{9} \frac{\overline{C}\, \overline{W}_0}
{C (T+\overline{T})^{\frac{p}{3} + n_X} } \tilde{X} \overline{\tilde{X}} \\
&& - \frac{n(p-3 n_X)}{9 C \overline{C}^{n-3} \Lambda^{n-3} (T+\overline{T})^{p/3}}
\tilde{X} \overline{\tilde{X}}^{n-1},
\nonumber\\
\delta F^T &=& 
\frac{p- n n_X}{p} \frac{ (T+\overline{T})^{1-p/3} }{C \overline{C}^{n-2} \Lambda^{n-3}}
\overline{\tilde{X}}^n+
\frac{p-3n_X}{3p} \frac{\overline{C}\, \overline{W}_0}
{ C^2 (T+\overline{T})^{p/3+n_X-1} } \tilde{X}{\tilde{X}},
\end{eqnarray}
and
\begin{equation}
F^{\tilde{X}} = - \frac{n(T+\overline{T})^{n_X-p/3}}{\Lambda^{n-3} \overline{C}^{n-3}}
(\overline{\tilde{X}})^{n-1}. \label{FX}
\end{equation}
Given that the perturbation to the scalar potential due to $\delta F^A$
is $\delta V = G_{A\bar{B}} \delta F^A \bar{F}^{\bar{B}}_0$, we can see that the leading order 
terms in powers of $\tilde{X}$ are ${\mathcal
O}\left(\tilde{X}^{2n-2}\right)$, ${\mathcal O}\left(m_{3/2}
\tilde{X}^n \right)$, and ${\mathcal O}\left(m_{3/2}^2 \tilde{X}^2
\right)$. Hence, we expect
\begin{eqnarray}
\tilde{X} &\sim& m_{3/2}^{\frac{1}{n-2}}, \nonumber \\
m_{3/2}^2 \tilde{X}^2 &\sim& m_{3/2} \tilde{X}^n \sim \tilde{X}^{2n-2}
\sim m_{3/2}^{\frac{2n-2}{n-2}}.
\end{eqnarray}
Ignoring the subleading order terms, which are suppressed by
$\tilde{X}/M_P$, $\tilde{X}/\Lambda$ (or $\Lambda/\tilde{X}$ for nonperturbative corrections) and $1/
\ln(M_{P} /m_{3/2})$, we find 
\begin{eqnarray}
\delta V &=& \left[ -\frac{1}{3} (p-3)(n-3) - (p- n n_X)\right]
\left[ \frac{F^C_0 \tilde{X}^n}{\Lambda^{n-3} C^{n-2} } + \mbox{h.c.} \right]
\nonumber \\
&& + \frac{n (p-3n_X) }{3} 
\left(\frac{F^C_0 \overline{\tilde{X}} \tilde{X}^{n-1} }{\Lambda^{n-3}
\overline{C} C^{n-3} } + \mbox{h.c.} \right) \nonumber \\
&& + \frac{n^2 (T+\overline{T})^{n_X - p/3}}{\Lambda^{2(n-3)}(C\overline{C})^{n-3}}
(\tilde{X} \overline{\tilde{X}} )^{n-1}.
\end{eqnarray}
Let us assume that $\tilde{X} = \overline{\tilde{X}}$ and $C = \overline{C}$, in which case we obtain
\begin{eqnarray}
\delta V = (n-3) \frac{ F^C_0 \tilde{X}^n + \overline{F}_0^C \overline{\tilde{X}}^n}
{\Lambda^{n-3} C^{n-2}} 
+ \frac{n^2 (T+\overline{T})^{n_X - p/3}}{\Lambda^{2(n-3)} C^{2(n-3)}} \left(
\tilde{X} \overline{\tilde{X}}\right)^{n-1}.
\end{eqnarray}
%Note that the final potential is independent of $p$.  
Minimizing $\delta V$ shows 
\begin{eqnarray}
\tilde{X} =
\left[ - \frac{n-3}{n(n-1)} \frac{\Lambda^{n-3} C^{n-3}}{(T+\overline{T})^
{n_X-p/3}}
\frac{F^{C*}}{C} \right]^{1/(n-2)}.
\end{eqnarray}
Using this result in  Eq.~(\ref{FX}) and comparing it with Eq.~(\ref{zerorelation}), 
we obtain
\begin{eqnarray}
\frac{F^{\tilde{X}}}{\tilde{X}} = \frac{n-3}{n-1} \frac{F^C}{C}.
\end{eqnarray}
Replacing $\tilde{X}$ with $X$ thus leads to the result
\begin{eqnarray}
\frac{F^X}{X} = -\frac{2}{n-1}\frac{F^C}{C}.
\label{higherordstab}
\end{eqnarray}
Since the modulus and anomaly contributions
are already comparable (see Eq.~(\ref{modulusanomaly}), this result indicates that all three contributions should be roughly equal for a very general class of superpotentials.  Interestingly, this result is also the same as that obtained in the case of deflected anomaly mediation  \cite{Pomarol:1999ie}, keeping in mind our definitions of $X$ and $\tilde{X}$.

In summary, we have demonstrated that in the absence of a bare superpotential mass term for $X$, both radiative stabilization and stabilization via higher-order superpotential self-interactions  lead to values of $F^X/X$ which are of the order of $F^C/C$, which result in comparable contributions to observable sector soft terms.  We now turn to the issue of computing the soft masses generated by these  F terms.

\section{MSSM Soft Supersymmetry Breaking Terms}
\label{susybrksect}
The F term vevs studied in the previous subsection control the nature of the soft terms induced by supersymmetry breaking. The specifics of the top-down model are no longer essential for our phenomenological purposes, and it suffices to simply take the F terms as parameters. By varying these parameters, we can obtain general mixtures of the moduli, gauge, and anomaly mediation.  With a given set of F terms, the soft mass spectrum can be computed using the spurion technique; it depend on the vevs of the modulus and compensator fields, the messenger scale, and the number of messengers, in a manner which we will now review.

Before we present the derivation of the soft terms, let us first recall the form of the MSSM superpotential.  In general, the superpotential of the MSSM fields $\Phi_i$ takes the form
\begin{equation}
W_{\rm MSSM}=\mu^0_{ij} \Phi_i\Phi_j +y^0_{ijk}\Phi_i\Phi_j\Phi_k,
\label{MSSMW}
\end{equation}
in which $\mu^0_{ij}$ are supersymmetric mass parameters, and $y^0_{ijk}$ are the (unnormalized) Yukawa couplings.  We assume that only trilinear couplings are present in the superpotential, and defer the discussion of supersymmetric mass terms ({\it i.e.} the $\mu$ problem of the MSSM) until later in the paper.  In addition to the couplings of Eq.~(\ref{MSSMW}), the $\Phi_i$ fields have K\"{a}hler metrics given by Eq.~(\ref{kahlermet}).   For the computation of the soft terms, the quantities
\begin{eqnarray}
Y_{i} = e^{ -K_0/3} Z_{i}
\end{eqnarray}
are useful for discussing the renormalization group evolution effects. The $Y_i$ are the coefficients of the bilinear terms in the nonholomorphic part of the superspace action: 
\begin{eqnarray}
\int d^4 \theta C \bar{C} G = \int d^4 \theta C \bar{C} \left( -3 \exp ( -K / 3 ) \right).
\end{eqnarray}
For our scenario, the $Y_i$ are given by 
\begin{equation}
Y_i =  \frac{1}{(T+\bar{T})^{n_i-\frac{p}{3}}}.
\end{equation}
To derive the observable sector soft terms, it is convenient to use the spurion technique, in which the 
couplings of the effective supergravity Lagrangian are regarded as functions in superspace, with the $\theta$-dependent parts of these couplings are generated by the F term vevs of the theory (for a review, see \cite{Giudice:1998bp}).  More precisely, in the presence of $F^C$, $F^T$ and $F^X$, the gauge kinetic function $f_a$ can be analytically continued to a 
superspace function
\begin{eqnarray}
f_a &=& \left. f_a \right|_0 
+ \theta^2 \left( 
 \left. \partial_C f_a
 % \frac{\partial f_a}{\partial C}
   \right|_0 F^C
+\left. \partial_T f_a
%\frac{\partial f_a}{\partial T}
 \right|_0 F^T + 
\left. \partial_X f_a
%\frac{\partial f_a}{\partial X} 
\right|_0 F^X 
\right), \label{spurionfa}
\end{eqnarray}
in which $|_0$ indicates that the value is taken at $\theta = \bar\theta = 0$. 
The K\"ahler metrics of the matter fields are also analytically continued 
into superspace as follows:
%For later purpose, we define
%\begin{eqnarray}
%\tilde{Y}_i = C \overline{C} Y_i  = -3 C \overline{C} \exp ( - K / 3 ). 
%\end{eqnarray}
\begin{eqnarray}
Y_i &=& \left. Y_i \right|_0 
+ \theta^2 
\left. \partial_{\Phi_A} Y_i
%\frac{\partial Y_i}{\partial \Phi^A} 
\right|_0 F^A
+ \bar{\theta}^2 \left. \partial_{\Phi_{\overline{B}}} Y_i
%\frac{\partial Y_i}{\partial \overline{\Phi}^{\overline{B}}}
\right|_0 \overline{F}^{\overline{B}}
+ \theta^2 \bar{\theta}^2 
%\sum_{A,\overline{B}}
\left.  \partial_{\Phi_A} \partial_{\Phi_{\overline{B}}}Y_i
%\frac{\partial^2 Y_i }{\partial \Phi^A \partial \overline{\Phi}^{\overline{B}}}
 \right|_0 F^A \overline{F}^{\overline{B}}, 
\label{spurionYi}
\end{eqnarray}
where once again, the indices $A,B$ denote $C$,$T$ and $X$. The superpotential Yukawa couplings in $y_{ijk}^0$  can in principle be a function of the moduli fields, in which case
\begin{eqnarray}
y^0_{ijk} = y^0_{ijk} |_0  + \theta^2 
\left. \partial_{\Phi_A} y^0_{ijk}
%\frac{\partial y_{ijk}}{\partial \Phi^A} 
\right|_0 F^A.
\label{spurionYukawa}
\end{eqnarray}
The MSSM soft supersymmetry breaking Lagrangian includes terms of the form
\begin{eqnarray}
\mathcal{L}_{\rm soft}=-m^2_i |\Phi^i|^2-\left [ \frac{1}{2} M_a \lambda^a \lambda^a + A_{ijk} y_{ijk} \Phi^i \Phi^j \Phi^k +\mbox{h.c.} \right ],
\end{eqnarray} 
in which $m^2_i$ are the soft scalar mass-squared parameters, $M_a$ are the gaugino masses, and $A_{ijk}$ are trilinear scalar interaction parameters.
%\begin{eqnarray}
%-m^2_i |\Phi^i|^2 
%\end{eqnarray}
%and trilinear scalar interaction terms
%\begin{eqnarray}
%-A_{ijk} y_{ijk} \Phi^i \Phi^j \Phi^k + \mbox{h.c.}.
%\end{eqnarray}
These terms are defined in the field basis in which the kinetic terms are canonically normalized; the physical Yukawa couplings $y_{ijk} = y^0_{ijk} / (Z_i Z_j Z_k)^{1/2}$ are in the 
definition of trilinear terms.  From the spurion couplings of Eqs.~(\ref{spurionfa})--(\ref{spurionYukawa}), 
%, (\ref{spurionYi}) and 
we obtain
\begin{eqnarray}
M_a &=& F^A \partial_A \log ({\rm Re\,}f_a),\label{Ma} \\
A_{ijk} &=& -F^A \partial_A \log \left( \frac{y^0_{ijk}}
{Y_i Y_j Y_k}\right), \nonumber \label{Aijk} \\
m_i^2 &=& - F^A \overline{F}^{\overline{B}} \partial_A \partial_{\overline{B}} \log
Y_i . \label{mi2}
\end{eqnarray}
Let us now consider the functional form of $f_a$, $Y_i$ and $y^0_{ijk}$.
First, note that the renormalization procedure gives an additional dependence of
these functions on the compensator field $C$. This effect arises because introducing a cut-off scale
$\Lambda_{\rm UV}$ explicitly breaks conformal invariance, such that  
to formally restore it, $\Lambda_{\rm UV}^2$ must be changed to $C \overline{C} \Lambda_{\rm UV}^2$. It can also be understood  by introducing Pauli-Villars regulators $\phi$, $\overline{\phi}$ with mass terms of the form
\begin{eqnarray}
 \int d^2 \theta C^3 \Lambda_{\rm UV} \phi \overline{\phi}.
\end{eqnarray}
With the redefinition $\tilde{\phi} = C \phi$, $\tilde{\phi}$ has an effective mass $C\Lambda_{\rm UV}$.
Therefore, the dependence on the renormalization scale $\mu$ is always accompanied by
\begin{eqnarray}
\frac{ C \overline{C} \Lambda^2_{\rm UV} }{\mu^2}.
\end{eqnarray}
Therefore, the conformal conpensator dependence of $f_a$ and $Y_i$  
can be extracted from 
\begin{eqnarray}
C \partial_C {\rm Re}~f_a 
&=& -\frac{1}{2} \mu \partial_\mu {\rm Re}f_a, \nonumber \\
C \partial_C Y_i
&=& -\frac{1}{2} \mu \partial_\mu Y_i. 
\end{eqnarray}
For the unnormalized Yukawa couplings $y^0_{ijk}$, there is no $C$ dependence due to the supersymmetric nonrenormalization theorem. Since $y^0_{ijk}$ is also assumed to be independent of $T$ and $X$, the expression for trilinear terms 
Eq.(\ref{Aijk}) can be 
reduced to 
\begin{eqnarray}
A_{ijk} = A_i + A_j + A_k, 
\end{eqnarray}
in which
\begin{eqnarray}
A_i = F^A \partial_A \log Y_i.
\end{eqnarray}
At the high scale $M_{\rm G}$, which we will take to be the GUT scale ($M_{\rm G}=2\times 10^{16}\,\mbox{GeV}$), 
\begin{eqnarray}
f_a (M_{\rm G}) &=& T^{l_a}, \nonumber \\
Y_i (M_{\rm G}) &=& \frac{1}{(T+\overline{T})^{n_i -\frac{p}{3} }},
\end{eqnarray}
In the above, $l_a=0,1$ depending on the type of D branes from which the gauge groups originate.   Since we wish to maintain gauge coupling unification at the GUT scale, we  assume that $l_a$ are the same for each of the SM gauge group factors.  To preserve gauge coupling unification, we also assume that the messenger pairs $\Psi$, $\overline{\Psi}$ are complete GUT multiplets e.g. of $SU(5)$, as is standard in many models of gauge mediation.  For general sets of messengers, it is useful to (re)define $N$ as
\begin{eqnarray} 
N = (\mbox{number of messenger pairs}) \times (\mbox{$SU(5)$ Dynkin index of
$\Psi$} ),
\end{eqnarray}
in which the $SU(5)$ Dynkin index for the fundamental representation ${\bf 5}$ is
normalized to unity.  Note that $N$ is the number of messenger pairs for the case in which $\Psi$ and $\overline{\Psi}$ are ${\bf 5}$ and ${\bf \overline{5}}$ representations; we will restrict ourselves to this situation in this work.

Above the mass scale of the messengers $M_{\rm mess}$, which is determined by the stabilization of $X$ by $M_{\rm mess}\equiv \langle X \rangle$, the matter content charged under the SM gauge group includes the MSSM fields and the messenger pairs. We will use a primed notation to denote the group 
theoretical parameters (e.g.~the $\beta$ functions) specific to this scale range ($M_{\rm mess}$ to
$M_{\rm G}$). The $\beta$-function coefficients $b^\prime_a$ are related to 
the MSSM beta function coefficients $b_a$ by 
\begin{eqnarray}
b^\prime_a = b_a + N,
\label{bprimedef}
\end{eqnarray}
in which 
\begin{eqnarray}
(b_3, b_2, b_1) = (-3, 1, \frac{33}{5}), 
\end{eqnarray}
for $SU(3)_C$, $SU(2)_L$ and $U(1)_Y$ (with GUT normalization, such that $b_1= 3/5 b_Y$).
\begin{figure}[t]
\begin{center}
{\epsfig{figure=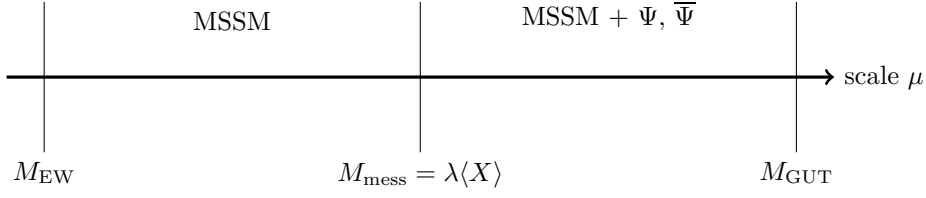}
\caption{The particle content of deflected mirage mediation as a function of scale. \label{scalefig}}}
\end{center}
\end{figure}
Below the messenger scale, only the MSSM matter fields are present as light degrees of freedom of the theory. For $\mu \ll M_{\rm mess}$, the gauge kinetic function can be easily obtained at 
one-loop: 
\begin{eqnarray}
{\rm Re~} f_a = {\rm Re~} f_a (M_{\rm G})
+\frac{b'_a}{16\pi^2} \ln \frac{M^2_{\rm G}}{X\overline{X}}
+\frac{b_a}{16\pi^2} \ln \frac{X \overline{X} C \overline{C}}{\mu^2}.
\label{faoneloop}
\end{eqnarray}
For the K\"ahler metric, the one-loop RG equation is not simply integrable: 
\begin{eqnarray}
\ln Y_i (\mu) &=& \ln Y_i (M_{\rm G}) + \frac{1}{16\pi^2} \sum_{j,k} 
\int^{\mu/\sqrt{C\overline{C}}}_{M_{\rm G}}
 \frac{d \mu}{\mu}
\left[ \frac{y^0_{ijk} \bar{y}^0_{ijk}}{Y_i Y_j Y_k}  \right] 
 \nonumber \\
&& + \frac{1}{16\pi^2} \sum_a
\int^{\mu/\sqrt{C\overline{C}}}_{M_{\rm G}}
 \frac{d\mu}{\mu} 
\left[ g_a^2 C_a(\Phi_i) \right].  \label{lnYi}
\end{eqnarray}
%CHECK YUKAWAS: y^0 or y?
The last term due to the gauge loop interaction can be explicitly integrated, 
but this is not possible for the Yukawa-dependent term. Fortunately, only the 
gauge-loop-induced term depends on $X$. We rewrite the last term of
Eq. (\ref{lnYi}) by
\begin{eqnarray}
-\sum_a \frac{2c_a}{b^\prime_a} \left( 
\ln \alpha_a^{-1} (M_{\rm G}) - \ln \alpha_a^{-1} (X) \right)
-\sum_a \frac{2c_a}{b_a} \left(
\ln \alpha_a^{-1} (X) - \ln \alpha_a^{-1} \left(\frac{\mu}{\sqrt{C\overline{C}}}
\right) 
\right),\nonumber \\
\end{eqnarray}
in which $\alpha_a$ is the conventionally defined gauge coupling parameter
$\alpha_a  =g_a^2/4\pi$, such that 
\begin{eqnarray}
\alpha^{-1}_a = 4 \pi\, {\rm Re~} f_a.
\end{eqnarray} 
The soft terms at the scale $\mu$ derived from Eqs. (\ref{faoneloop}) 
and (\ref{lnYi}) using Eqs.~(\ref{Ma}-\ref{mi2}) automatically 
include the renormalization group evolution of the soft parameters. This implies that we obtain the
correct soft mass formulae once we identify the contributions from the  threshold scale and then apply the usual RG equations for the soft mass parameters. One caveat is that above $M_{\rm mess}$, it is necessary to take into account the presence of the messenger fields in the beta functions.   Although Eq.~(\ref{lnYi}) is an integral equation, we can easily see that the Yukawa term does not
contribute to the threshold correction at $M_{\rm mess}$ by comparing 
its value infinitesimally above and below the messenger scale.  \\

\noindent We now present the MSSM soft terms for deflected mirage mediation, including the values at the GUT scale $M_{\rm G}$ and the messenger threshold effects at $M_{\rm mess}$:
\begin{itemize}
\item{\bf Gaugino Masses.} The gaugino mass parameters are given by
\begin{eqnarray}
  M_a (M_{\rm G}) &=& \frac{F^T}{T+\overline{T}}
  + \frac{g_0^2}{16\pi^2}  b'_a\frac{F^C}{C} \\ \label{softgaugino}
  M_a (M_{\rm mess}^-)& =& M_a(M_{\rm mess}^+)
+ \Delta M_a,
\end{eqnarray}
in which the threshold corrections are
\begin{eqnarray}
  \Delta M_a = - N\frac{g_a^2(M_{\rm mess})}{16\pi^2} \left( \frac{F^C}{C}
  + \frac{F^X}{X} \right).
   \label{softgauginothresh}
\end{eqnarray}
In the above, $g_0$ is the unified gauge coupling at $M_{\rm G}$, and $b'_a$ are given by Eq.~(\ref{bprimedef}) (our convention is that $b'_a<0$ for asymptotically free theories).

\item{\bf  Trilinear terms.}
Recalling that
\begin{eqnarray}
A_{ijk} = A_i+ A_j + A_k,
\end{eqnarray}
we have
\begin{eqnarray}
A_i (\mu = M_{\rm G}) &=& \left (\frac{p}{3}-n_i \right) \frac{F^T}{T+\overline{T}}
- \frac{\gamma_i}{16\pi^2} \frac{F^C}{C} ,  \label{softA}
\end{eqnarray}
in which $\gamma_i$ denotes the anomalous dimension of $\Phi_i$. Note that there are no
threshold contributions to the trilinear terms due to the messengers.  
\item{\bf Soft scalar masses. }
The scalar mass-squared parameters are given by
\begin{eqnarray}
m_i^2 (\mu = M_{\rm G}) = \left (\frac{p}{3}-n_i \right) \left|\frac{F^T}{T+\overline{T}}\right|^2
%\nonumber \\
%&& 
-\frac{\theta'_i}{32 \pi^2} \left( \frac{F^T}{T+\overline{T}}\frac{F^{\overline{C}}}{\overline{C}} + \mbox{h.c.} \right) \nonumber \
\label{softscalar}
 - \frac{\dot{\gamma}'_i}{(16\pi^2)^2} \left|\frac{F^C}{C}\right|^2,
%\nonumber \\
\end{eqnarray}
\begin{eqnarray}
m_i^2 (\mu = M_{\rm mess}^-) &=& m_i^2 (\mu = M_{\rm mess}^+)
+ \Delta m_i^2,
\end{eqnarray}
where the threshold corrections are
\begin{eqnarray}
\Delta m_i^2 & = &\sum_a 2 c_a N
\frac{g_a^4 (M_{\rm mess}) }{(16\pi^2)^2}
 \left( \left|\frac{F^X}{X}\right|^2 + \left|\frac{F^C}{C}
\right|^2
+ \frac{F^X}{X}
%{M_{\rm mess}
 \frac{F^{\overline{C}}}{\overline{C}} +\mbox{h.c.} \right), \label{softscalarthresh}
\end{eqnarray}
For completeness, $\gamma_i$, $\dot{\gamma}_i$, $\theta_i$ and their primed counterparts are given in Appendix A.
\end{itemize}

\section{Renormalization Group Analysis and Low Energy Mass Spectra}
\label{phenosect}
In this section, we will show the renormalization group (RG) running and the low energy mass spectra for the supersymmetry breaking parameters of Eqs.~(\ref{softgaugino})--(\ref{softscalarthresh}).

In what follows, we will now restrict ourselves to the case in which the K\"{a}hler potential of $T$ is of the ``no-scale" form ($p=3$).  For the MSSM superpotential, we assume that R-parity is conserved and bare superpotential mass terms are absent, such that the only terms are the standard Yukawa couplings required for fermion mass generation.  Here we do not explicitly address the $\mu/B\mu$ problem of the MSSM, except to say that in a given string theory model, supersymmetric bare mass terms are expected to be absent for the light ({\it i.e.}, below string scale) fields of the observable sector.   However, gauge mediated models have a well-known $\mu/B\mu$ problem.  Furthermore, due to the  anomaly mediated terms, the Giudice-Masiero mechanism for solving the $\mu$ problem \cite{Giudice:1988yz} results in a $B$ term of the order of the gravitino mass, which leads to a fine-tuning problem.  The best option for addressing this issue may then be the addition of extra singlets \cite{Kim:1983dt}; for a recent discussion see \cite{Nakamura:2008ey}.

We also comment here on the issue of flavor and CP violation in this scenario.  As previously stated, we assumed a diagonal K\"{a}hler metric for the MSSM fields $\Phi_i$ (see Eq.~(\ref{fullkahler}))  to avoid flavor-changing neutral currents.  However, it is well known that flavor-violating effects can also be induced from renormalization group running if these diagonal terms are not universal, which would correspond to generation-dependent modular weights.  Here we will assume generation-independent modular weights for simplicity, although the most stringent bounds occur for the first two generations.   Even with this form of the leading order terms, higher-order corrections K\"{a}hler potential couplings may exhibit nontrivial flavor violation; a more comprehensive study of this issue is found in \cite{Choi:2008hn}.  

Turning to effects of CP violation, an inspection of  Eqs.~(\ref{softgaugino})--(\ref{softscalarthresh}) clearly demonstrates that if there are nontrivial relative phases between $F^T/(T+\overline{T})$, $F^C/C$, and $F^X/X$, there will generically be irremovable flavor-independent CP-violating phases in the soft supersymmetry breaking terms.  Generically, if these phases are $O(1)$, the electron and neutrino electric dipole moments will exceed experimental bounds unless the superpartner masses are in the multi-TeV range.  In this paper, we will assume that the F terms are all real, though a thorough exploration of CP violation in this context may be worthwhile.

With these assumptions, it is useful to express the anomaly and gauge mediated contributions in terms of the modulus mediated contribution, which is governed by $F^T/(T+\overline{T})\equiv m_0$ (see Eq.~(\ref{graveq})).   Following  the mirage mediation literature, we define $\alpha_m$ as 
\begin{equation}
\frac{F^C}{C}=\alpha_m\ln \frac{M_P}{m_{3/2}} \frac{F^T}{T+\overline{T}}= \alpha_m\ln \frac{M_P}{m_{3/2}} m_0.
\end{equation}
Our parameter $\alpha_m$ is the $\alpha$ of mirage mediation.  To account for the gauge mediated terms, we will also define $\alpha_g$ by
\begin{equation}
\frac{F^X}{X}=\alpha_g \frac{F^C}{C}=\alpha_g\alpha_m \ln\frac{M_P}{m_{3/2}} m_0.
\end{equation}
With these definitions, $m_0$ sets the overall mass scale of the soft supersymmetry breaking terms, and the dimensionless parameters $\alpha_m$ and $\alpha_g$ denote the relative importance of anomaly mediation and gauge mediation, respectively.  With this parametrization, the soft terms at $M_G$ take the form
\begin{eqnarray}
\label{gaugino1}
M_a(M_{\rm G})&=&m_0\left [1+\frac{g_0^2}{16\pi^2}b_a^\prime \alpha_m \ln \frac{M_P}{m_{3/2}}\right ] ,\\
\label{trilinear1}
A_i(M_{\rm G})&=&m_0 \left [(1-n_i)-\frac{\gamma_i}{16\pi^2} \alpha_m \ln \frac{M_P}{m_{3/2}}\right ] ,\\
\label{mssq1}
m_i^2(M_{\rm G})&=&m_0 ^2 \left [(1-n_i)-\frac{\theta'_i}{16 \pi^2} \alpha_m \ln \frac{M_P}{m_{3/2}} -\frac{\dot{\gamma}'_i}{(16\pi^2)^2}\left (\alpha_m \ln \frac{M_P}{m_{3/2}}\right )^2 \right ],
\end{eqnarray}
and the threshold terms are given by
\begin{eqnarray}
\label{threshgaug1}
\Delta M_a&=& -m_0 N\frac{g_a^2(M_{\rm mess})}{16\pi^2}   \alpha_m \left (1 +\alpha_g \right ) \ln \frac{M_P}{m_{3/2}} ,\\
\label{threshmssq1}
\Delta m_i^2&=&m_0^2\sum_a 2 c_a N
\frac{g_a^4 (M_{\rm mess}) }{(16\pi^2)^2} \left [\alpha_m  (1+\alpha_g)  \ln \frac{M_P}{m_{3/2}}\right ]^2.
\end{eqnarray}
In principle, $\alpha_m$ and $\alpha_g$ can be considered to be continuous parameters, although in specific string-motivated models they are typically given by discrete values.  In this paper, we will consider specific (discrete) string-motivated values of $\alpha_m$ and $\alpha_g$, and defer an analysis of more general possibilities for future work.   To focus on the new features which arise from the gauge mediated terms, here we will fix $\alpha_m=1$, which is the value predicted in the standard KKLT model, and focus on the $\alpha_g$ values which result from the stabilization of the mediator field $X$ via the stabilization mechanisms discussed in Section~\ref{theorysect}. Note that in the radiative stabilization case $\alpha_g=-1$ (see Eq.~(\ref{radstab})).  For stabilization via higher-order superpotential terms, $\alpha_g$ can take on different discrete values depending on the details and dynamical origin of the higher-order terms (see Eq.~(\ref{higherordstab})).  

\subsection{(Deflected) Mirage Unification}
%\noindent $\bullet$ {\bf Mirage Unification and the Mirage Mediation Limit}.
Before turning to a more detailed analysis of specific parameter sets, we will first address the issue of mirage unification.  In the KKLT mirage mediation scenario, one of the most distinctive features of the soft terms is the unification of the gaugino masses at the mirage unification scale $M_{\rm mir}$:
% which is given by
\begin{equation}
M_{\rm mirage}=M_{\rm G}\left (\frac{m_{3/2}}{M_P} \right )^\frac{\alpha_m}{2}.
\label{mirageunif}
\end{equation}
Eq.~(\ref{mirageunif}) indicates that $\alpha_m=2$ is needed to have mirage unification at a TeV scale, which is desirable from the point of view of electroweak scale fine tuning.  However, it is well known that this value of $\alpha_m$ is not easily obtained within the KKLT approach.  Another intriguing feature of mirage mediation is that the soft trilinear scalar couplings (the $A$ terms) and the soft scalar mass-squares can also unify at $M_{\rm mirage}$.  Whether mirage unification happens for these parameters depends on the modular weights ({\it i.e.}, if $\sum_{l=i,j,k}(1-n_l)=1$ for fields with nonvanishing Yukawa couplings $y_{ijk}$), or on whether the effects of the Yukawas on the running are negligible, as is the case for the first and second generations; further details on this issue can be found in \cite{Choi:2005uz}.  

In deflected mirage mediation, we find a similar mirage unification phenomenon for the gaugino masses.  From the form of the soft terms of Eq.~(\ref{gaugino1}) and Eq.~(\ref{threshgaug1}), the new mirage unification scale for the gauginos (see also \cite{Everett:2008qy}) is
\begin{equation}
M_{\rm mirage}=M_{\rm G}\left (\frac{m_{3/2}}{M_P} \right )^\frac{\alpha_m\rho }{2},
\label{deflmirageunif}
\end{equation}
in which $\rho$ is given by
\begin{equation}
\rho= \frac{1+ \frac{2Ng_0^2}{16\pi^2}  \ln \frac{M_{\rm GUT}}{M_{\rm mess}}}{1- \frac{\alpha_{\rm m} \alpha_{\rm g} Ng_0^2}{16\pi^2} 
 \ln \frac{M_P}{m_{3/2}}}.
% \right )^{\alpha_{\rm m} \alpha_{\rm g}}}.
\label{rhodef}
\end{equation}
The mirage unification scale of the gauginos is thus deflected from the mirage mediation result.  The size of the deflection is dependent on $\alpha_g$, $N$, and $M_{\rm mess}$, which govern the size of the messenger thresholds.  As we will demonstrate with specific examples, the deflected mirage gaugino mass unification scale can be as low as the TeV scale, even for $\alpha_m=1$.

For the $A$ terms and the soft scalar mass-squares, the mirage unification behavior no longer happens in general in the presence of the messengers.  The exception is when the messenger scale is below the scale of mirage unification which would occur in the absence of the messenger thresholds, since the theory is then effectively the same as mirage mediation below $M_{\rm mess}$.   While the soft scalar mass-squares of the light generations typically no longer unify, they can display a quasi-conformal behavior ({\it i.e.}, the masses do not vary with scale) below $M_{\rm mess}$ in certain cases.  Examples which display these features will be shown later in the paper (additional examples can be found in \cite{Everett:2008qy}).

\begin{figure}
\centering
\subfigure[ ]{\includegraphics[width=6cm] 
{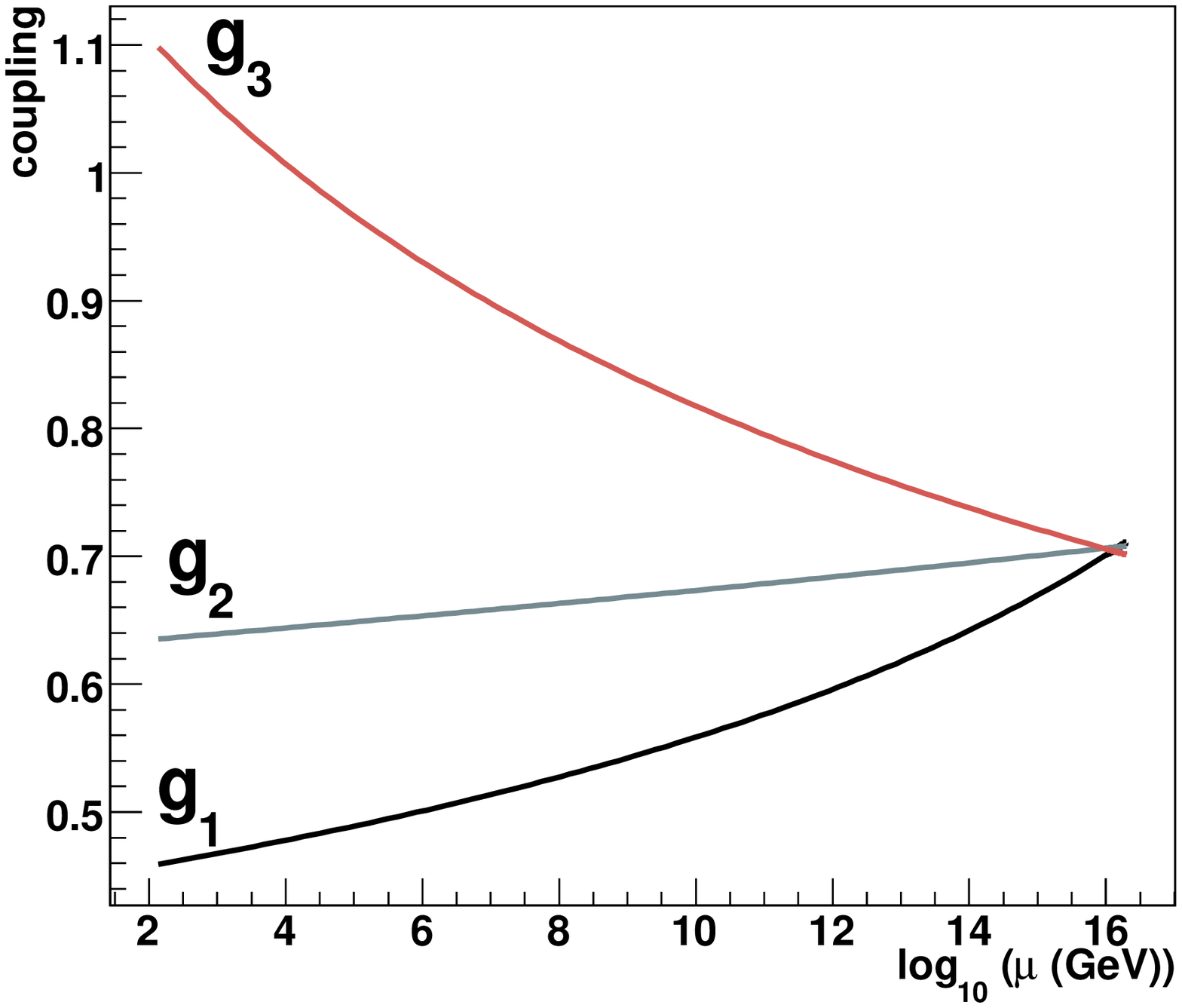}}
\qquad
\subfigure[]{\includegraphics[width=6cm] 
{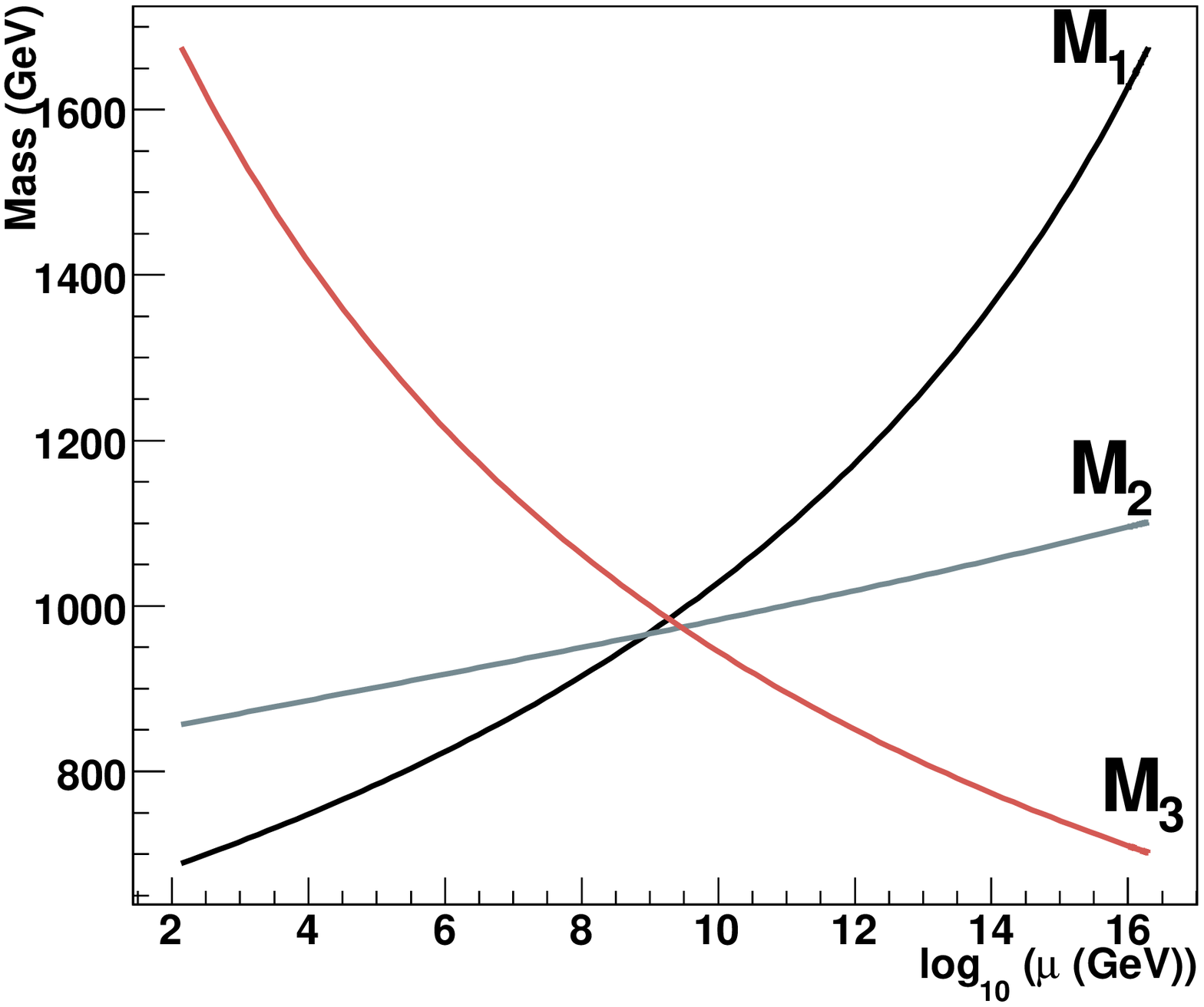}}\\
\subfigure[ ]{\includegraphics[width=6cm] 
{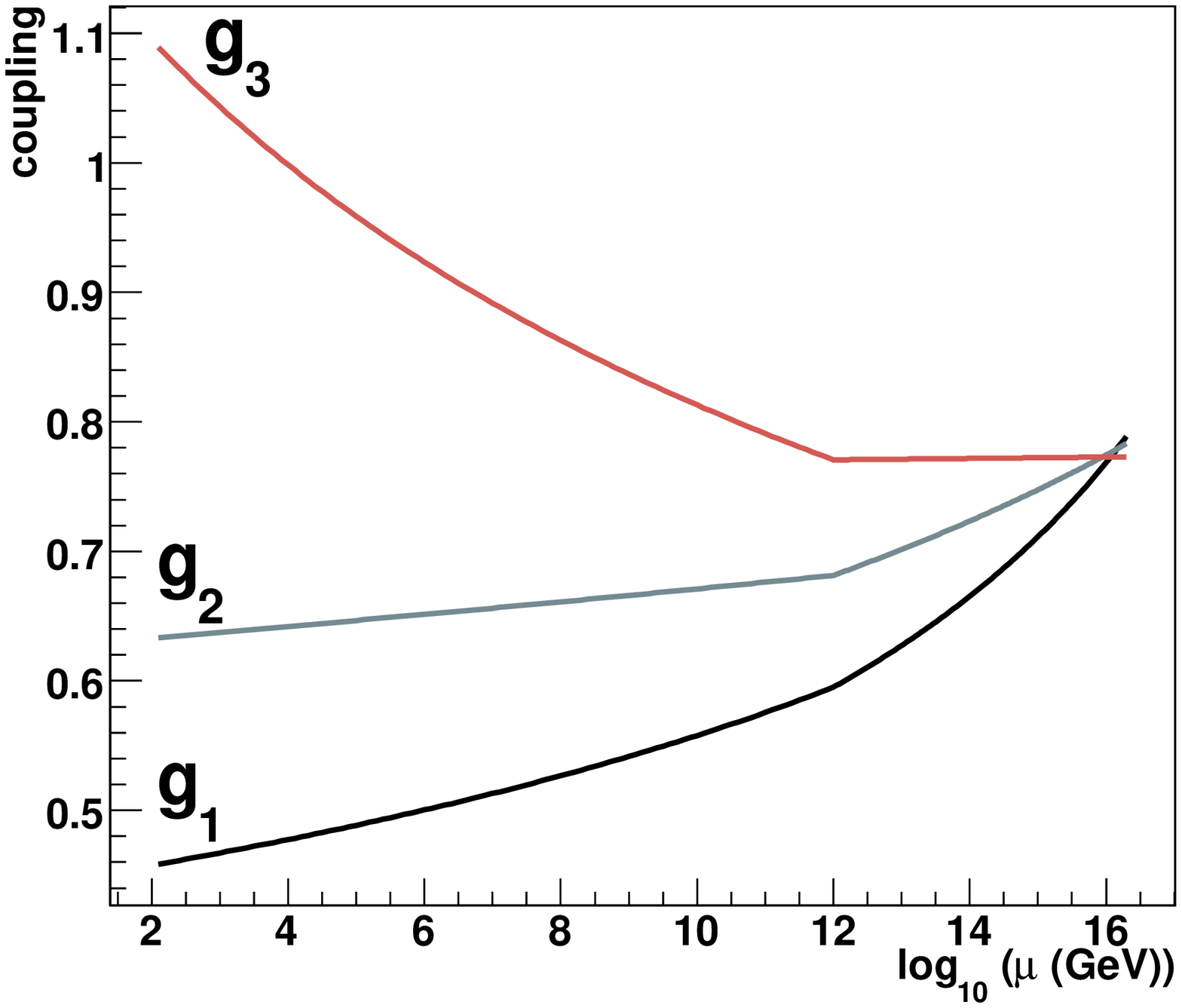}}
\qquad
\subfigure[]{\includegraphics[width=6cm] 
{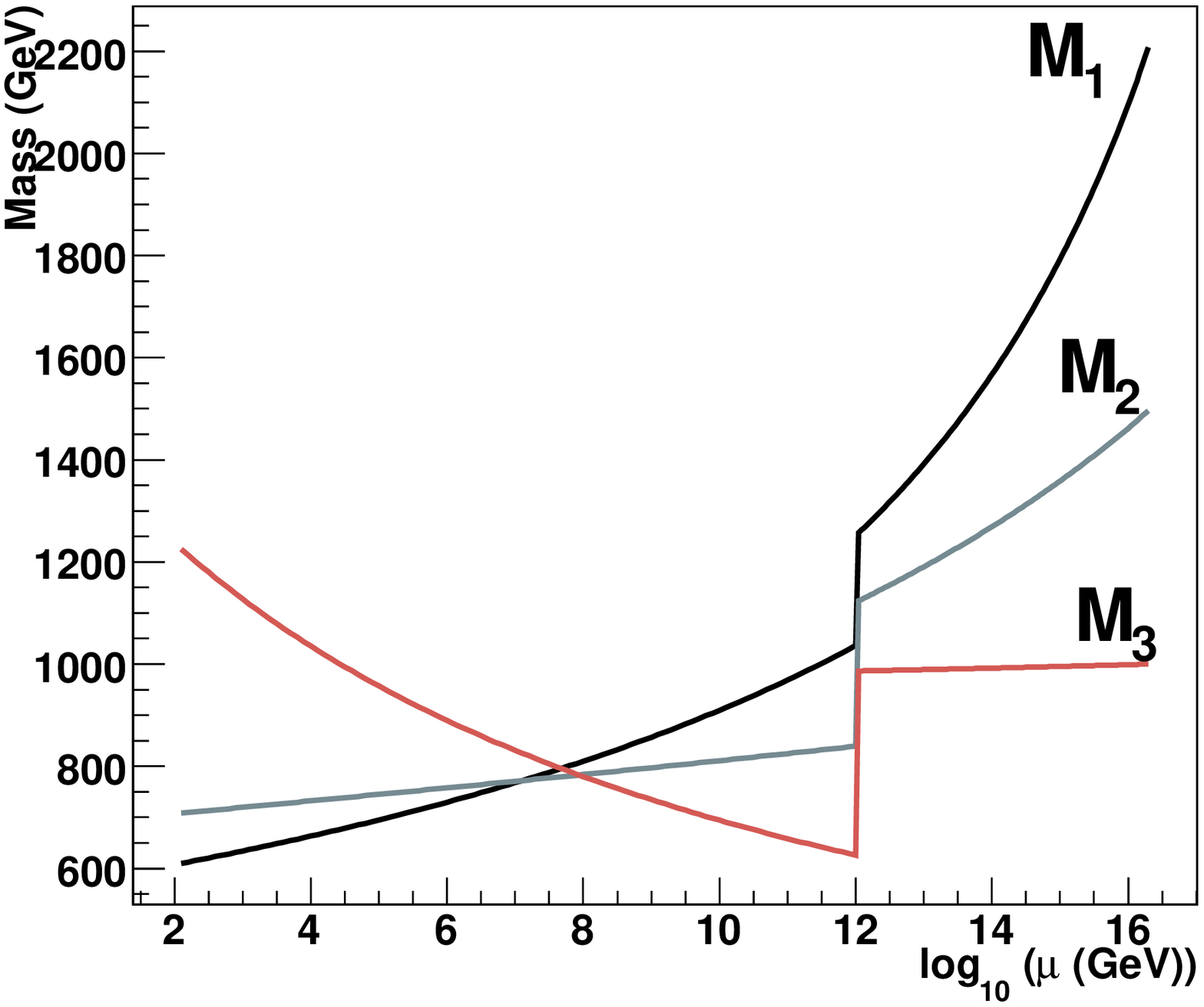}}
\caption{The renormalization group evolution of the gauge couplings and gaugino masses, in which $\alpha_m=1$, $\alpha_g=0$, $M_{\rm mess}=10^{12}\,\mbox{GeV}$, $m_0=1\,\mbox{TeV}$, and (a)--(b) $N=0$, and (c)--(d) $N=3$, in which case there are messenger threshold effects.  The standard KKLT model is obtained for $N=0$.  \label{gauginomirage}}
\end{figure}

In Eqs.~(\ref{deflmirageunif})--(\ref{rhodef}), the mirage mediation result of Eq.~(\ref{mirageunif}) is obtained only if $N=0$.  This demonstrates that the mirage mediation limit is not reached when gauge mediation is switched off ($\alpha_g\rightarrow 0$); it only occurs when the messengers are removed from the theory at all scales ($N=0$).  The reason is that the messengers affect the MSSM beta functions above the messenger scale, which in turn affects the anomaly mediated terms.  To show this feature explicitly (which can also be seen from the expressions for the soft terms), in Fig.~\ref{gauginomirage} we show the gauge coupling and gaugino mass renormalization group evolution in two models with $\alpha_m=1$, $\alpha_g=0$, $m_0=1\,\mbox{TeV}$ and (a)--(b) $N=0$, which is the standard KKLT model, and (c)--(d) $N=3$.  For $N=0$, mirage unification occurs at the usual scale of $\sim 10^9\,\mbox{GeV}$ (for $\alpha_m=1$), while for $N=3$ the mirage unification scale is deflected to a lower value (below $10^8\,\mbox{GeV}$).

\subsection{Examples}
In addition to $\alpha_m$ and $\alpha_g$, there are both continuous and discrete parameters of the model.  The continuous parameters are $m_0$, $\tan\beta$, and the messenger scale $M_{\rm mess}$; the discrete parameters are the number of messengers $N$, the modular weights $n_i$, the exponents $l_a$ in the gauge kinetic functions, and the sign of $\mu$ (we trade $\mu$ and $B$ for $m_Z$ and $\tan\beta$).  Here we will consider several values for $M_{\rm mess}$ and fix the other parameters as follows:
\begin{eqnarray}
&&n_Q = n_U = n_D = n_L = n_E = \frac{1}{2},\;\; n_{H_u} = n_{H_d} = 1,\;\;l_1=l_2=l_3=1, \nonumber \\
&&m_0=\mbox{1 TeV},\;\;\tan\beta=10,\;\;N=3,\;\; \mbox{sign} (\mu) = +1.
\label{paramchoice}
\end{eqnarray}
This choice of modular weights is often used in the mirage mediation literature, though nonuniversal modular weights may also be of interest.  In this paper, we focus on the effects from the messenger thresholds, and leave the issue of general $n_i$ for future exploration.

We will now describe the renormalization group evolution and the resulting low energy superpartner mass spectrum for sample parameter sets as a function of $\alpha_g$ and $M_{\rm mess}$ (for $\alpha_m=1$), with the other parameters fixed by Eq.~(\ref{paramchoice}).  These models are not specifically chosen to obtain certain desired low energy features, such as a value of the neutralino relic density within the allowed experimental range.  It is straightforward to scan the parameter space and find such allowed points; two examples of such dark-matter allowed parameter sets were previously presented in \cite{Everett:2008qy}.  Here the goal is to exhibit the $\alpha_g$ dependence and to examine the implications of theoretically motivated values of $\alpha_g$ on the gaugino mirage unification scale and the pattern of soft masses at low energies.\\

\noindent $\bullet$ {\bf Example 1: Radiative Stabilization.} 
First, we will consider the case in which $X$ has no superpotential self-interaction terms, and instead is stabilized by radiative supersymmetry breaking effects.  As shown in Eq.~(\ref{radstab}), $F^X/X=-F^C/C$, and hence $\alpha_g=-1$.  It can immediately be seen from Eqs.~(\ref{softgauginothresh})--(\ref{softscalarthresh}) (or equivalently Eqs.~(\ref{threshgaug1})--(\ref{threshmssq1})) that the threshold effects identically vanish, and hence the only effect of the messengers is to change the beta functions above the messenger scale.    

\begin{figure}[t]
\centering
\subfigure[ Gaugino masses.]{\includegraphics[width=6cm] 
{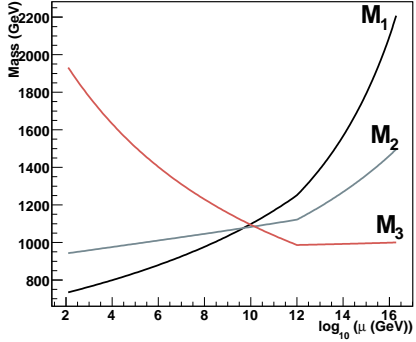}}
\qquad
\subfigure[First family soft scalar masses. ]{\includegraphics[width=6cm] 
{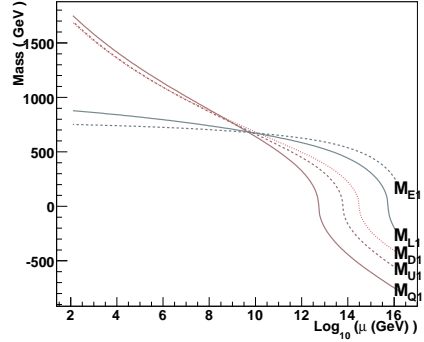}}\\
\subfigure[Third family soft scalar masses. ]{\includegraphics[width=6cm] 
{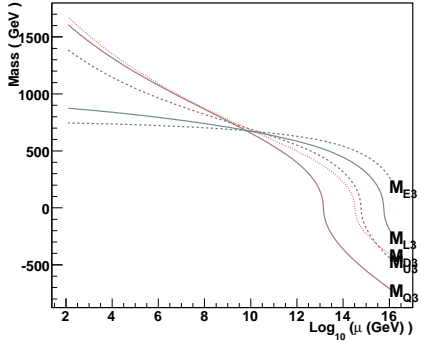}}
\qquad
\subfigure[Third family soft trilinear terms. ]{\includegraphics[width=6cm] 
{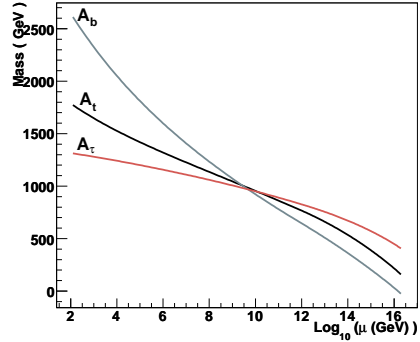}}
\caption{The renormalization group evolution of (a) the gaugino masses, (b) the first family soft scalar mass-squares, (c) the third family soft scalar mass-squares, and (d) the third generation $A$ terms, for the case of radiative stabilization, with $\alpha_m=1$, $\alpha_g=-1$, and $M_{\rm mess}=10^{12}\,\mbox{GeV}$. \label{radstabfig1}}
\end{figure}

To see the resulting effects on the low energy values of the soft masses, in Fig.~\ref{radstabfig1} we show the renormalization group evolution of (a) the gaugino masses, (b)--(c) the soft scalar masses of the first and third generations, and (d) the third generation trilinear scalar terms, in a model with $M_{\rm mess}=10^{12}\,\mbox{GeV}$.
Fig.~\ref{radstabfig1} shows that the resulting physics is then similar to that of mirage mediation: the gaugino masses unify at a high scale, and the $A$ terms and the soft scalar masses show an  approximate mirage unification behavior (due to our choice of modular weights in Eq.~(\ref{paramchoice})).  The subsequent renormalization group evolution drives the gluino to be rather heavy; this heavy gluino controls the renormalization group flow for many of the other soft terms.  The sleptons, whose RG equations do not depend on the gluino mass at one loop, tend to remain light, while the squarks are driven to be relatively heavy.  The $A$ terms can also be relatively large, which can lead to a relatively light stop due to left-right splitting.  The gauginos have too large a mass hierarchy to mix strongly, so the lightest supersymmetric partner (LSP) is nearly pure bino.  Bino annihilation tends to give too large a relic density, but when $\tilde{\tau}$ or $\tilde{t}$ is light enough coannhilation effects can reduce the relic density to an acceptable value.  With this set of parameters, the relic density remains too large since coannihilation effects are not significant; however, it is possible to adjust other parameters to obtain a low energy spectrum consistent with dark matter constraints.   

\begin{figure}
\centering
\subfigure[ Gaugino masses.]{\includegraphics[width=6cm] 
{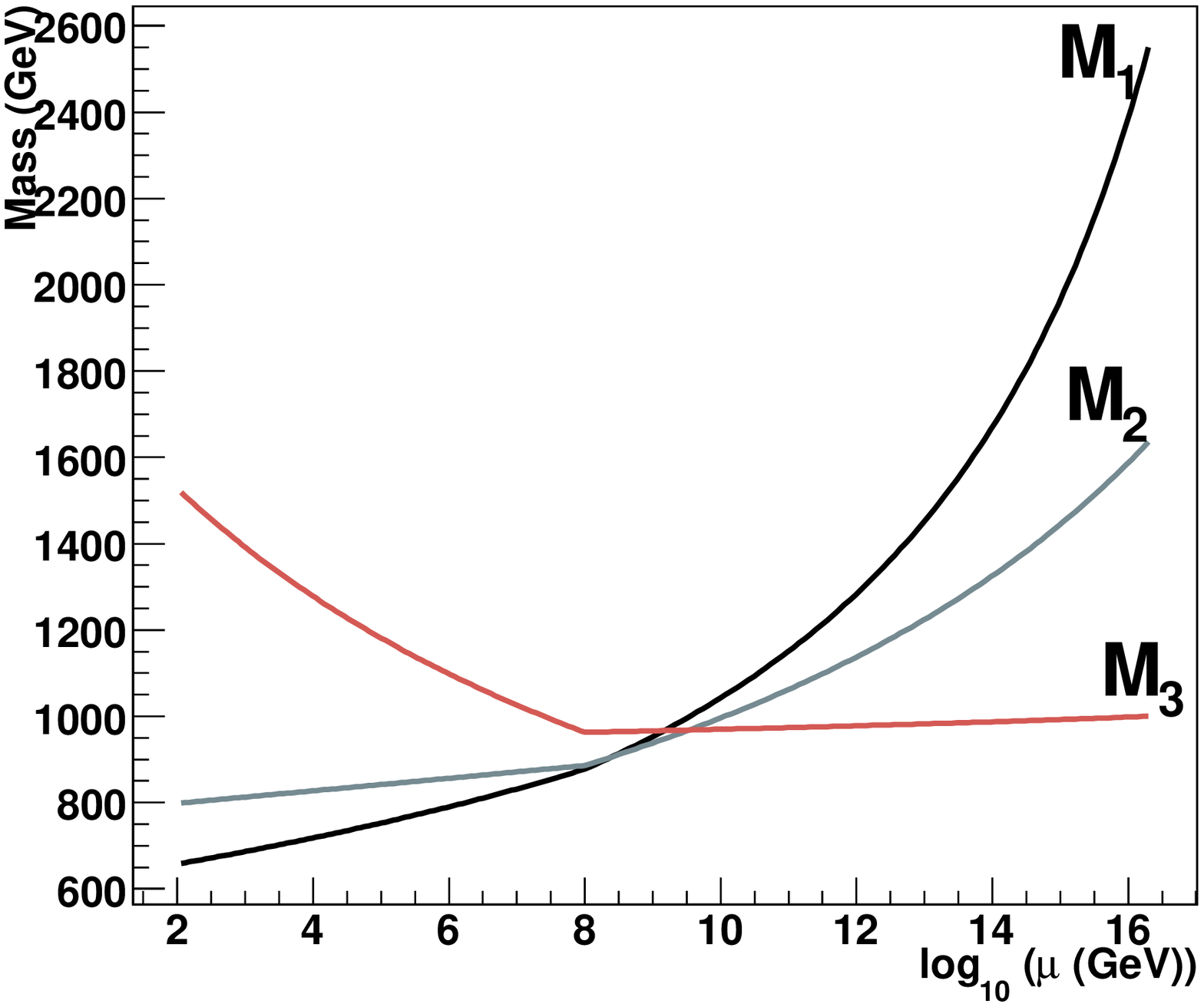}}
\qquad
\subfigure[Third family soft scalar masses. ]{\includegraphics[width=6cm] 
{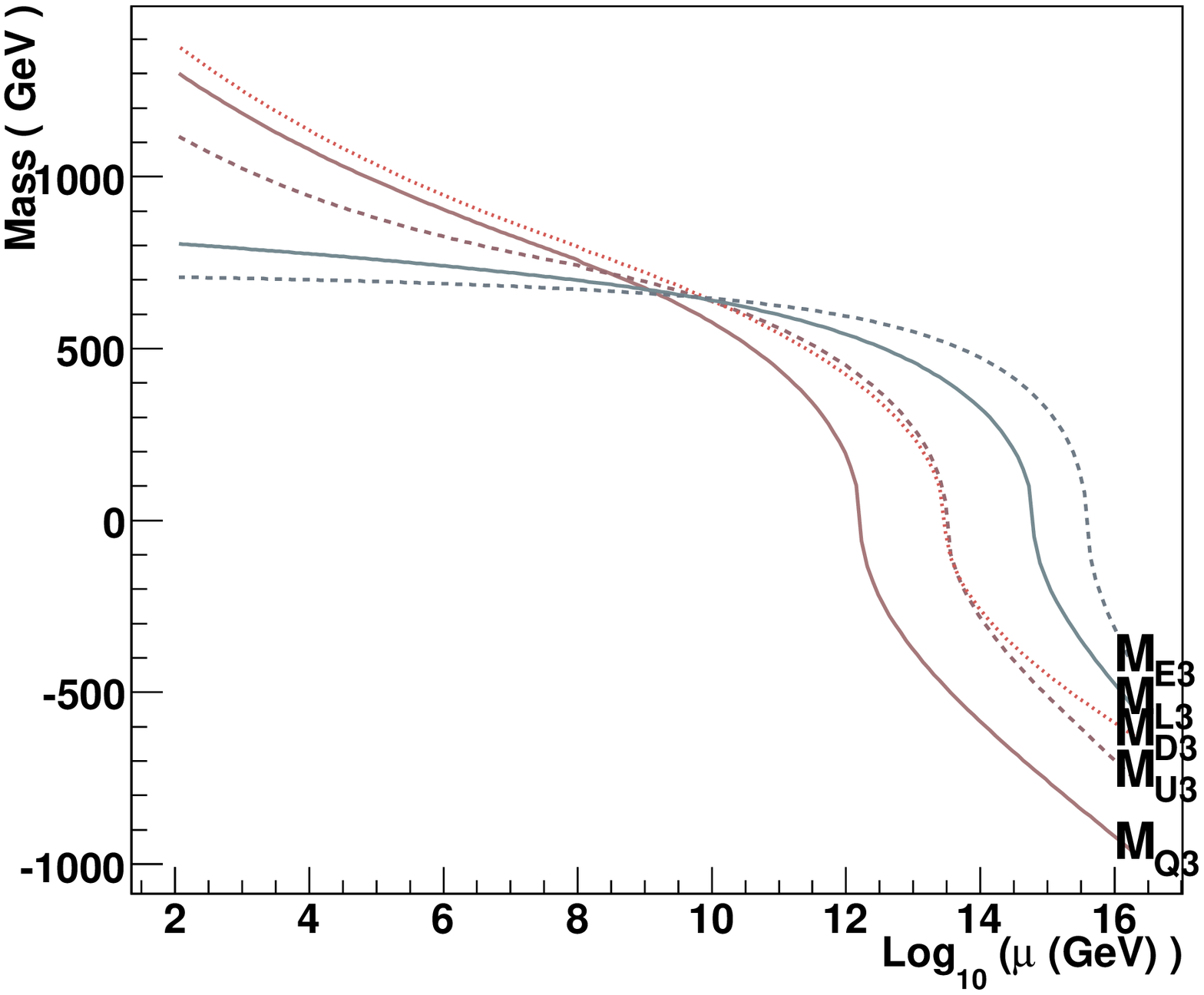}}
\caption{The renormalization group evolution of (a) the gaugino masses and the (b) the third family soft scalar mass-squares for $\alpha_m=1$, $\alpha_g=-1$, and $M_{\rm mess}=10^{8}\,\mbox{GeV}$. \label{radstabfig2}}
\end{figure}

If the messenger scale is lowered, for example to  $M_{\rm mess}=10^8\,\mbox{GeV}$, the mirage unification behavior for the gauginos occurs before the decoupling of the messengers; the $A$ terms and scalar mass-squares also unify at the mirage scale.  In Fig.~\ref{radstabfig2}, we show the RG running of (a) the gauginos and (b) the third family soft scalar mass-squares.  Since the presence of the messengers flattens the running of the gluino (in fact, $b_3'=0$ for $N=3$), the mirage unification behavior occurs at a lower scale than in the case of a higher $M_{\rm mess}$.  The resulting slightly lighter gluino results in a lighter SUSY mass spectrum.  The lightest superpartner is again almost purely bino, and due to the lighter superpartners, the relic abundance is larger than it was in the previous case with $M_{\rm mess}=10^{12}\,\mbox{GeV}$. \\

\noindent $\bullet$ {\bf Example 2: Stabilization via Nonrenormalizable Operators}.  For our second example, we will consider a model in which the $X$ stabilization occurs through superpotential couplings of the form given in Eq.~(\ref{higherxint}).  As shown in Eq.~(\ref{higherordstab}),  this leads to $F^X/X=(-2/(n-1))F^C/C$, such that $\alpha_g=-2/(n-1)$.   We will consider the case of $n=4$, corresponding to stabilization through a perturbative nonrenormalizable operator, which implies $\alpha_g=-2/3$.  With this stabilization mechanism, $\langle X\rangle \sim (\Lambda m_{3/2})^{1/2}$, and hence if $\Lambda\sim M_P$, then $M_{\rm mess}\sim 10^{10}\,\mbox{GeV}$.  In this example, we will set $\alpha_m=1$, $\alpha_g=-2/3$, and consider two possible values of the messenger scale: $M_{\rm mess}=10^{10}\,\mbox{GeV}$ and $M_{\rm mess}=10^{6}\,\mbox{GeV}$, which corresponds to a smaller value of the cutoff scale $\Lambda$. 

\begin{figure}
\centering
\subfigure[ Gaugino masses.]{\includegraphics[width=6cm] 
{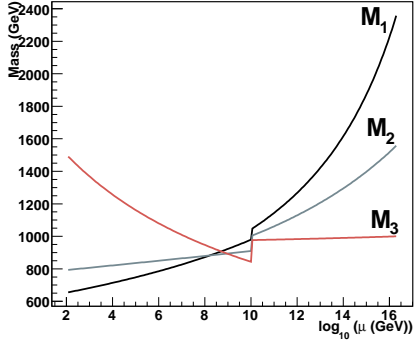}}
\qquad
\subfigure[First family soft scalar masses. ]{\includegraphics[width=6cm] 
{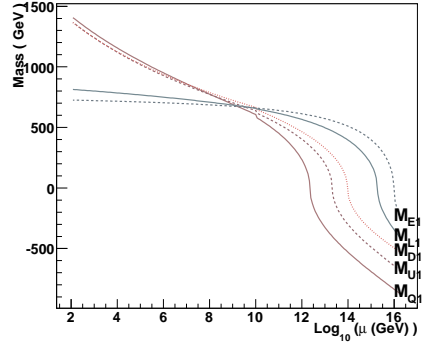}}\\
\subfigure[ Gaugino masses.]{\includegraphics[width=6cm] 
{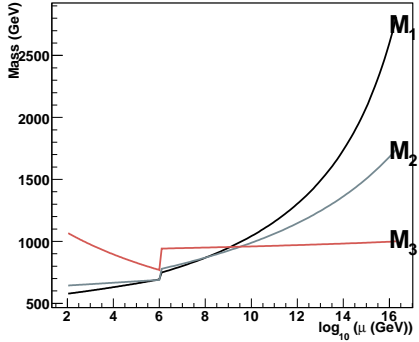}}
\qquad
\subfigure[First family soft scalar masses. ]{\includegraphics[width=6cm] 
{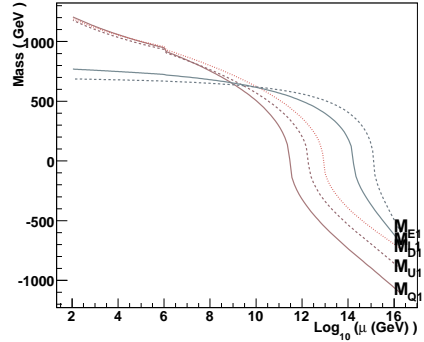}}
\caption{The renormalization group evolution of (a) the gaugino masses and (b) the first generation soft scalar mass-squares for $M_{\rm mess}=10^{10}\,\mbox{GeV}$, and (c) the gaugino masses and (d) the first family scalar mass-squares for $M_{\rm mess}=10^{6}\,\mbox{GeV}$, for the case of the stabilization of $X$ by a nonrenormalizable superpotential term, with $\alpha_m=1$ and $\alpha_g=-2/3$.\label{highordstab1}}
\end{figure}

In Fig.~\ref{highordstab1}, we present the renormalization group evolution of the gaugino masses and the soft scalar mass-squares of the first generation for $M_{\rm mess}=10^{10}\,\mbox{GeV}$ in panels (a) and (b), and for $M_{\rm mess}=10^{6}\,\mbox{GeV}$ in panels (c) and (d).
The physics of this case resembles that of the radiative stabilization model, since $\alpha_g<0$.  For the case with a higher messenger scale, the small threshold effects which result when $\alpha_g<0$ again lead to mirage unification of the gauginos and soft scalar masses at a relatively high scale of about $10^9\,\mbox{GeV}$ (the scale dependence the soft mass-squares and $A$ terms of the third generation are not displayed explicitly, as they too resemble that of the $\alpha_g=-1$ example of Fig.~\ref{radstabfig1} and Fig.~\ref{radstabfig2}).  The gluino is relatively heavy, which again leads to relatively large masses for the superpartners with $SU(3)$ charges and a large stop mass splitting.  The LSP is once again almost pure bino, with a characteristically too large relic abundance.  When the messenger scale is smaller, the gluino is significantly lighter, resulting in a more compressed superpartner spectrum.  However, the gauginos mix more strongly due to the smaller mass splitting of $M_1$ and $M_2$.  Though the LSP remains almost pure bino, the slight increase in the wino component acts to lessen the relic density; this competes with effects of the lighter spectrum, which act to increase the relic abundance.  For this particular parameter set, we find that while the neutralino relic abundance is too large to be allowed by dark matter constraints, it is slightly smaller for $M_{\rm mess}=10^6\,\mbox{GeV}$ than it is for $M_{\rm mess}=10^{10}\,\mbox{GeV}$.

The $n=4$ case corresponds to the lowest nonrenormalizable operator that results in the stabilization of $X$ (note that the renormalizable $n=3$ case results in $\alpha_g=-1$, just as in the case of radiative stabilization).  More generally, the mechanism of stabilizing $X$ by nonrenormalizable operators leads to $-1<\alpha_g<0$, with higher powers of $n$ resulting in $\alpha_g$ approaching its limiting value of zero.  For higher-order operators, clearly $\alpha_g$, while still negative, has a smaller magnitude, such that the cancellation of the messenger threshold effects is less efficient and more dramatic effects can appear.  In our previous paper \cite{Everett:2008qy}, we presented a sample model with similar features, with $\alpha_m=1$ and $\alpha_g=-1/2$ ({\it i.e.}, $n=5$), in which the dark matter abundance was in the allowed range due to stop coannihilation. \\

\noindent $\bullet$ {\bf Example 3: Stabilization via Nonperturbative Effects}.  For our last example, we consider models in which that the superpotential self-interaction of the $X$ field originates from nonperturbative dynamics, such that $n<0$ in Eq.~(\ref{higherxint}).  From Eq.~(\ref{higherordstab}), we see that this will result in $\alpha_g>0$.   To obtain a concrete example, we will choose $n=-1$, in which case $\alpha_g=1$.  Since $\langle X \rangle $ is now given by $\sim (\Lambda^4/m_{3/2})^{1/3}$,  if $m_{3/2}\sim 30\,\mbox{TeV}$ and $\Lambda\sim 10^{10}\,\mbox{GeV}$, then $M_{\rm mess}\sim 10^{12}\,\mbox{GeV}$.  As in previous examples, we will consider two values of the messenger scale: $M_{\rm mess}=10^{12}\,\mbox{GeV}$ and $M_{\rm mess}= 10^{8}\,\mbox{GeV}$.

\begin{figure}[t]
\centering
\subfigure[ Gaugino masses.]{\includegraphics[width=6cm] 
{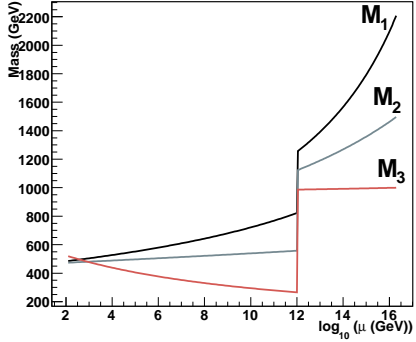}}
\qquad
\subfigure[First family soft scalar masses. ]{\includegraphics[width=6cm] 
{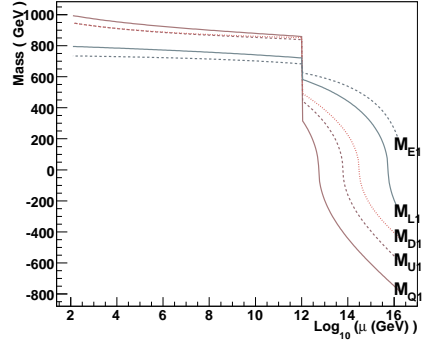}}\\
\subfigure[Third family soft scalar masses. ]{\includegraphics[width=6cm] 
{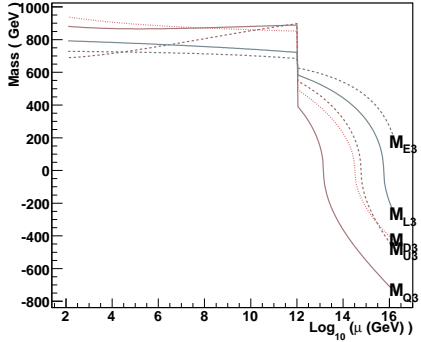}}
\qquad
\subfigure[Third family soft trilinear terms. ]{\includegraphics[width=6cm] 
{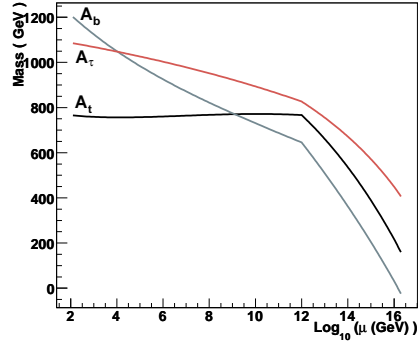}}
\caption{The renormalization group evolution of (a) the gaugino masses, (b) the first family soft scalar mass-squares, (c) the third family soft scalar mass-squares, and (d) the third generation $A$ terms, for the case of stabilization due to nonperturbative superpotential terms, with $\alpha_m=1$, $\alpha_g=1$, and $M_{\rm mess}=10^{12}\,\mbox{GeV}$. \label{nonpertfig1}}
\end{figure}
In Fig.~\ref{nonpertfig1}, the renormalization group running of (a) the gaugino masses, (b) the first generation soft scalar mass-squares, (c) the third generation soft scalar masses, and (d) the third generation trilinear scalar terms, are shown for $\alpha_m=\alpha_g=1$ and $M_{\rm mess}=10^{12}\,\mbox{GeV}$.
Here we can immediately see dramatically different features emerge for the case of $\alpha_g>0$.  In particular, the threshold effects from gauge mediation become important and can drive the gauginos to unify at a low scale of order 1 TeV. The gauginos become nearly degenerate and mix strongly with each other; accordingly, the lightest neutralino is a general mixture between the bino, wino, and higgsinos.  In this example, the
relic density falls within the experimental limits, indicating that the LSP is indeed a ``well-tempered neutralino" \cite{ArkaniHamed:2006mb}.   (We note in passing, however, that this parameter choice is still not suitable for benchmark studies, because the lightest Higgs mass is too low to be allowed by experimental bounds.)

Since the messenger threshold corrections drive the gaugino masses, and particularly the gluino mass,  
to be lighter, the low energy soft masses correspond in most cases to a  ``compressed'' SUSY spectrum (see e.g. \cite{compressed,gaugemessenger}).  Since  the gluino is light, the squarks also remain relatively light, though significantly heavier than the gluino.  The very light gluino leads to an intriguing feature of the RG flow of the soft scalar masses, which is the quasi-conformal fixed point behavior at scales below the messenger scale.  This feature can be seen clearly for the lighter generations in panel (b), and to a lesser extent for the third generation in panel (c) due to Yukawa couplings, of Fig.~\ref{nonpertfig1}.  The messenger thresholds lower the gluino mass to such an extent that it does not strongly participate in the RG evolution of the other soft parameters, leading to the quasi-conformal behavior for the scalar sparticles (particularly the squarks).  Furthermore, the $A$ terms no longer unify at a high mirage scale, since $M_{\rm mess}$ is larger than the original mirage unification scale.

\begin{figure}[t] %******************
\centering
\subfigure[ Gaugino masses.]{\includegraphics[width=6cm] 
{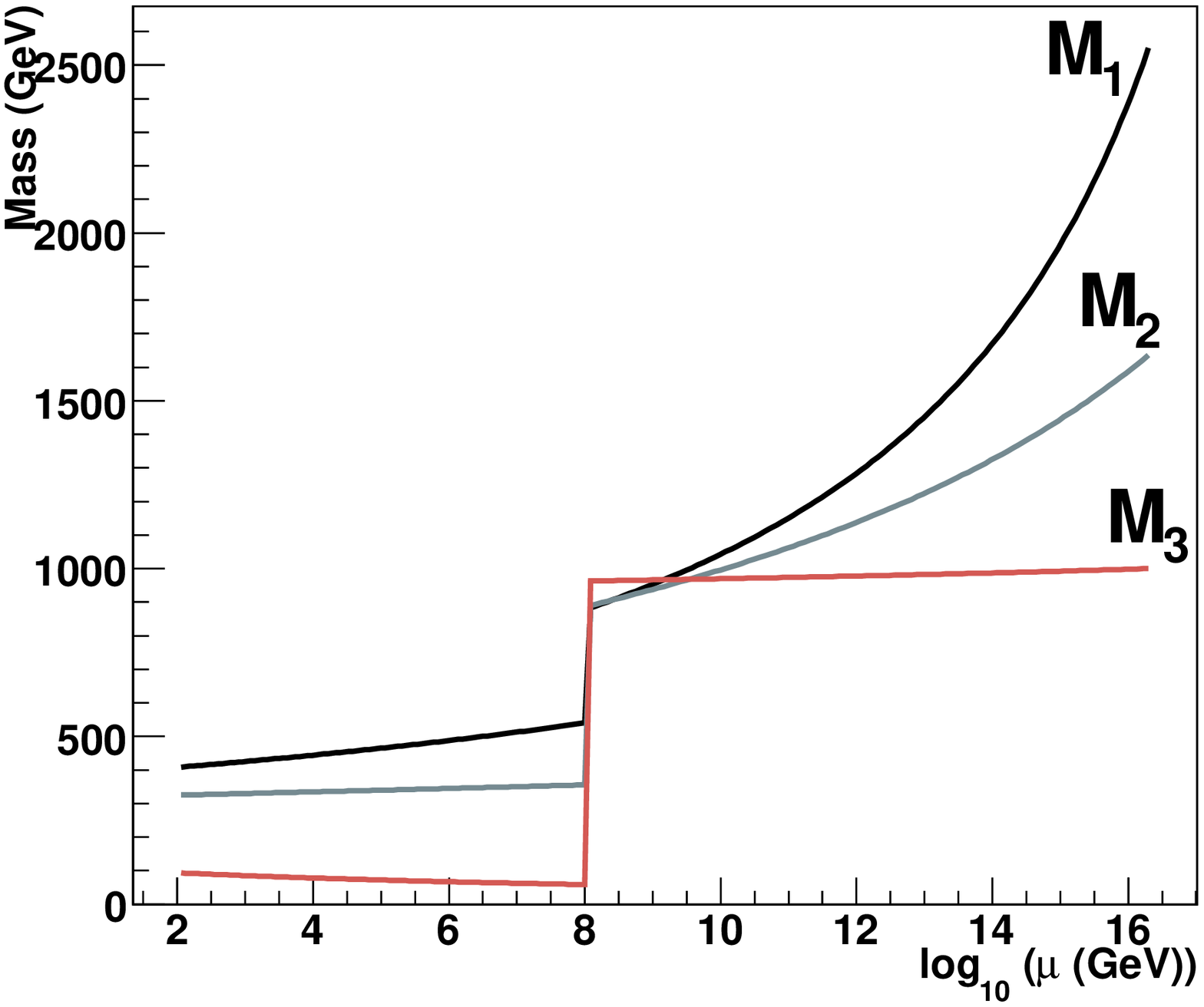}}
\qquad
\subfigure[First family soft scalar masses. ]{\includegraphics[width=6cm] 
{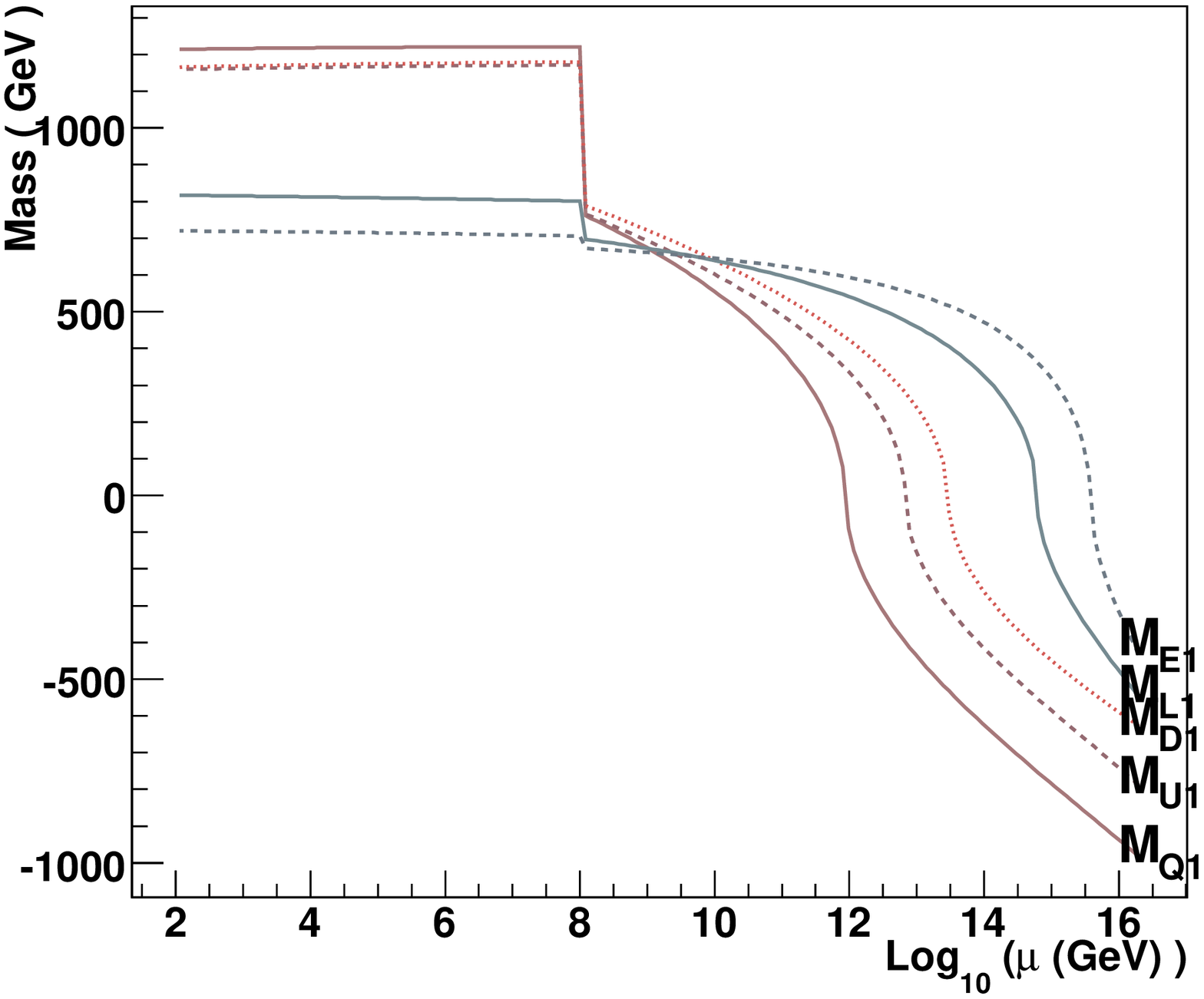}}\\
\subfigure[Third family soft scalar masses. ]{\includegraphics[width=6cm] 
{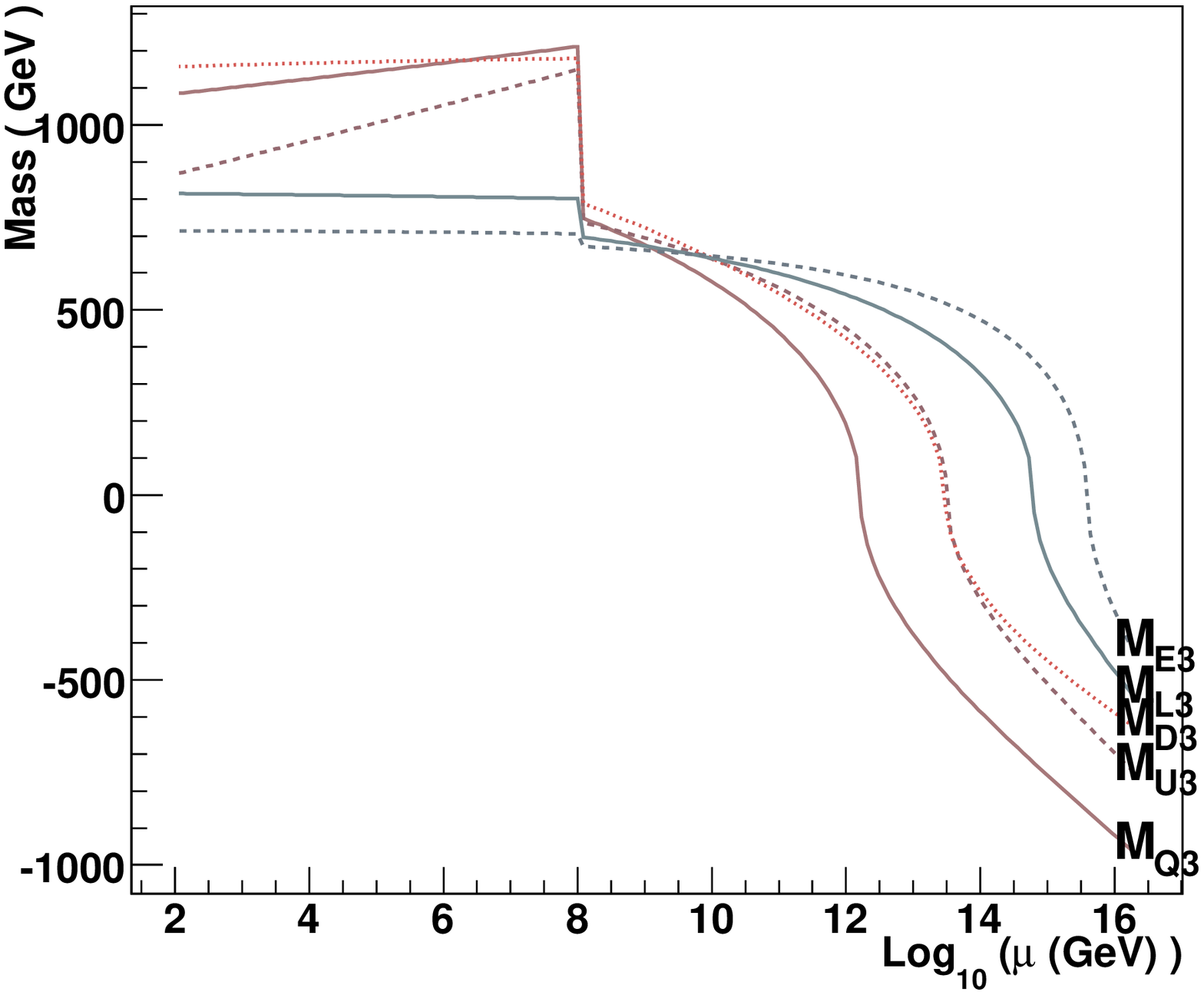}}
\qquad
\subfigure[Third family soft trilinear terms. ]{\includegraphics[width=6cm] 
{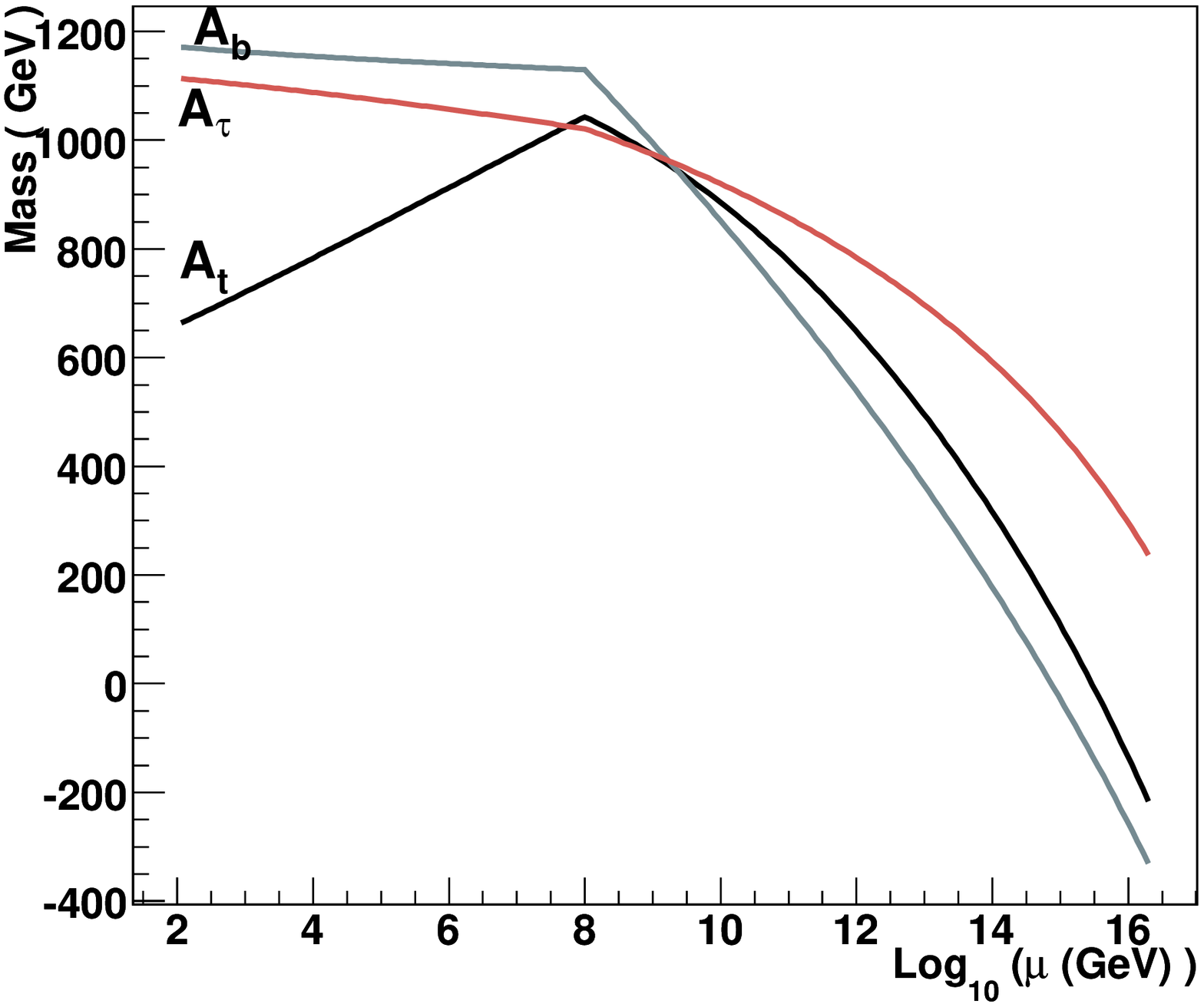}}
\caption{The renormalization group evolution of (a) the gaugino masses, (b) the first family soft scalar mass-squares, (c) the third family soft scalar mass-squares, and (d) the third generation $A$ terms, for the case of stabilization due to nonperturbative superpotential terms, with $\alpha_m=1$, $\alpha_g=1$, and $M_{\rm mess}=10^{8}\,\mbox{GeV}$. \label{nonpertfig2}}
\end{figure}

Several of the qualitative features shown above change in the case in which the messenger scale is lowered to $M_{\rm mess}=10^{8}\,\mbox{GeV}$, as shown in Fig.~\ref{nonpertfig2}.
In this case, we see that the messenger scale is below the mirage unification scale, and hence the gauginos, soft scalar mass-squares, and the $A$ terms all unify at a high scale of order $10^9\,\mbox{GeV}$.  The threshold effects drive the gluino mass to very small values, such that the gluino is the LSP.  This results again in a quasi-conformal behavior of the soft mass-squares.  The superpartner spectrum is slightly less compressed, since the point at which the gluino becomes very light occurs at a later point in the running.  The spectrum includes (i) very light states, which are the gluino LSP, two light neutralinos (including the wino-dominated next lightest superpartner (NLSP)), and the lighter chargino, (ii) moderately heavy states ($\sim 800\,\mbox{GeV}$) which include the sleptons,  charginos, and neutralinos, and (iii) heavier squarks, of order $1.1-1.2\,\mbox{TeV}$.   For $\alpha_g>0$, the threshold effects can either lead to a gluino LSP or a slepton LSP, if the RG running above the messenger threshold is not strong enough to drive to drive the binos and winos sufficiently light.

\section{Conclusions}
\label{concl}
In this paper, we have investigated both theoretical and phenomenological aspects of deflected mirage mediation, which is a recently proposed string-motivated model of supersymmetry breaking in which modulus (gravity) mediation, gauge mediation, and anomaly mediation all contribute to the soft terms.  In deflected mirage mediation, the minimal KKLT mirage mediation model, which predicts comparable modulus and anomaly mediated terms, is extended to include gauge mediated terms.  The gauge mediated terms arise from additional visible sector fields: the gauge singlet mediator field $X$ and $N$ vectorlike pairs of messenger fields with SM charges.  The mediator $X$ acquires a SUSY breaking F term through supergravity effects due to its mixing with the K\"{a}hler modulus $T$.  

We have explicitly demonstrated that $X$ can be stabilized either by radiative corrections or higher-order terms in the superpotential, which can be either nonrenormalizable or nonperturbative in origin.  The resulting gauge mediated contributions to the observable sector soft terms are proportional to the anomaly mediated contributions, with the proportionality factor $\alpha_g$ (with $F^X/X=\alpha_g F^C/C$).   Since the anomaly mediated and modulus mediated terms in this scenario are comparable (their ratio is given by $\alpha_m$, which is the $\alpha$ of mirage mediation), all three contributions are relevant for the soft terms.  We computed the MSSM soft terms and investigated the renormalization group evolution and resulting pattern of superpartner masses at low energies for the minimal KKLT model  ($\alpha_m=1$), focusing on values of $\alpha_g$ that emerge from each of the $X$ stabilization mechanisms.  We find that the gaugino mass mirage unification scale is deflected from the value obtained in the absence of the messengers.  The low energy physics strongly depends on the sign of $\alpha_g$.  For $\alpha_g<0$, which corresponds to stabilization by radiative effects or perturbative nonrenormalizable superpotential terms, the threshold effects are small and the basic pattern of soft masses is similar that of mirage mediation, with a bino-dominated LSP.  For $\alpha_g>0$, which corresponds to stabilization by nonperturbatively-generated operators, the threshold effects can be large;  this can deflect the gaugino mass mirage unification scale to TeV values, and typically drives the stops and gluinos to be light. In this case, the LSP can even be the gluino, though it can also be a mixed bino-wino-higgsino ({\it i.e.}, ``well-tempered") neutralino which is allowed by dark matter constraints.

In conclusion, deflected mirage mediation provides a rich setting in which to investigate the theoretical and phenomenological implications of low energy supersymmetry.  One particular direction of interest is to concrete model building would be of interest.  Deflected mirage mediation also has the advantage that it encompasses three of the standard supersymmetry breaking mediation mechanisms within one generalized framework, in which the relative contributions from each mechanism can be adjusted by fixing a small number of parameters.  Although here we considered discrete parameter values motivated from the top-down approach, we can take a bottom-up approach in which these parameters are taken to be continuous, which allows us to dial between many of the different known models of supersymmetry breaking.  The resulting generalized framework lends itself to more general studies of the phenomenological implications of low energy supersymmetry at the LHC.  Work along these lines is currently in progress \cite{uslong2}.

\acknowledgments
This work is supported by the U.S. Department of Energy grant DE-FG-02-95ER40896.  P.O. is also supported in part by the NSF Career Award PHY-0348093 and a Cottrell Scholar Award from Research Corporation. 

\appendix

\section{Anomalous dimensions}
At one loop, the anomalous dimension is given by 
\begin{eqnarray}
\gamma_i = 2 \sum_a g_a^2 c_a(\Phi_i) - \frac{1}{2}\sum_{lm} |y_{ilm}|^2,
\label{gammaexp}
\end{eqnarray}
in which $c_a$ is the quadratic Casimir, and $y_{ilm}$ are the normalized Yukawa couplings.  Here we will consider only the Yukawa couplings of the third generation $y_t$, $y_b$, and $y_\tau$.  For the MSSM fields $Q$, $U^c$, $D^c$, $L$, $E^c$, $H_u$ and $H_d$,
the anomalous dimensions are
\begin{eqnarray}
\gamma_{Q,i} &=& \frac{8}{3} g_3^2 + \frac{3}{2} g_2^2 + \frac{1}{30} g_1^2
- (y_t^2 + y_b^2) \delta_{i3},
\nonumber \\
\gamma_{U,i} &=& \frac{8}{3} g_3^2 + \frac{8}{15} g_1^2
- 2 y_t^2 \delta_{i3},
\nonumber \\
\gamma_{D,i} &=& \frac{8}{3} g_3^2 + \frac{2}{15} g_1^2
- 2 y_b^2 \delta_{i3},
\nonumber \\
\gamma_{L,i} &=& \frac{3}{2} g_2^2 + \frac{3}{10} g_1^2
-y_\tau^2 \delta_{i3},
\nonumber \\
\gamma_{E,i} &=& \frac{6}{5} g_1^2
-2 y_\tau^2 \delta_{i3},
\nonumber \\
\gamma_{H_u} &=& \frac{3}{2} g_2^2 + \frac{3}{10} g_1^2
-3 y_t^2 \nonumber \\
\gamma_{H_d} &=& \frac{3}{2} g_2^2 + \frac{3}{10} g_1^2
- 3 y_b^2 - y_{\tau}^2,
\end{eqnarray}
respectively.
Above $M_{\rm mess}$, the beta function of the gauge couplings
changes because of the messenger fields.  However, $\gamma_i$ does not change according to
Eq.~(\ref{gammaexp}), and hence $\gamma'_i = \gamma_i$.  The $\dot{\gamma}_i$'s are given by the expression
\begin{eqnarray}
\dot{\gamma}_i=2\sum_a g_a^4b_a c_a(\Phi_i) - \sum_{lm} |y_{ilm}|^2b_{y_{ilm}},
\end{eqnarray}
in which $b_{y_{ilm}}$ is the beta function for the Yukawa coupling $y_{ilm}$.  For the MSSM fields, the 
$\dot{\gamma}_i$'s are given by
\begin{eqnarray}
\dot\gamma_{Q,i}
&=& \frac{8}{3} b_3 g_3^4 + \frac{3}{2} b_2 g_2^4 + \frac{1}{30} b_1 g_1^4
- (y_t^2 b_t + y_b^2 b_b ) \delta_{i3} \nonumber \\
\dot\gamma_{U,i}
&=& \frac{8}{3} b_3 g_3^4 + \frac{8}{15} b_1 g_1^4
- 2 y^4_t b_t \delta_{i3} \nonumber \\
\dot\gamma_{D,i}
&=& \frac{8}{3} b_3 g_3^4 + \frac{2}{15} b_1 g_1^4
- 2 y^4_b b_b \delta_{i3} \nonumber \\
\dot\gamma_{L,i}
&=& \frac{3}{2} b_2 g_2^4 + \frac{3}{10} b_1 g_1^4
 - y_\tau^2 b_\tau \delta_{i3} \nonumber \\
\dot\gamma_{E,i}
&=& \frac{6}{5} b_1 g_1^4
 - 2 y_\tau^2 b_\tau \delta_{i3} \nonumber \\
\dot\gamma_{H_u}
&=& \frac{3}{2} b_2 g_2^4 + \frac{3}{10} b_1 g_1^4
 - 3 y^2_t b_t \nonumber \\
\dot\gamma_{H_d}
&=& \frac{3}{2} b_2 g_2^4 + \frac{3}{10} b_1 g_1^4
 - 3 y_b^2 b_b - y^2_\tau b_\tau,  \label{dotgammaexp}
\end{eqnarray}
where
\begin{eqnarray}
b_t &=& 6 y_t^2 + y_b^2 -\frac{16}{3} g_3^2 - 3 g_2^2 - \frac{13}{15} g_1^2,
\nonumber \\
b_b &=& y_t^2 + 6 y_b^2 + y_\tau^2 -\frac{16}{3} g_3^2 - 3 g_2^2
- \frac{7}{16} g_1^2, \\
b_\tau &=& 3 y_b^2 + 4 y_\tau^2 - 3 g_2^2 -\frac{9}{5} g_1^2.
\end{eqnarray}
$\dot\gamma^\prime_i$ is obtained by replacing $b_a$ with $b'_a = b_a + N$
in Eq.~(\ref{dotgammaexp}). Finally, ${\theta_i}$, which appears in the mixed modulus-anomaly term in the soft scalar mass-squared parameters, is given by
\begin{eqnarray}
\theta_i = 4 \sum_a g_a^2 c_a(Q_i) - \sum_{i,j,k} |y_{ijk}|^2
( p- n_i -n_j- n_k).
\end{eqnarray}
For the MSSM fields, the ${\theta_i}$ are
\begin{eqnarray}
\theta_{Q,i} &=& \frac{16}{3} g_3^2 + 3 g_2^2 + \frac{1}{15} g_1^2
-2 ( y_t^2 (p-n_{H_u}-n_Q -n_U) + y_b^2 (p-n_{H_d} -n_Q - n_D )) \delta_{i3},
\nonumber \\
\theta_{U,i} &=& \frac{16}{3} g_3^2 + \frac{16}{15} g_1^2
- 4 y_t^2 (p-n_{H_u} - n_Q - n_U ) \delta_{i3}, \nonumber \\
\theta_{D,i} &=& \frac{16}{3} g_3^2 + \frac{4}{15} g_1^2
- 4 y_b^2 (p- n_{H_d} - n_Q - n_D ) \delta_{i3}, \nonumber \\
\theta_{L,i} &=& 3 g_2^2 + \frac{3}{5} g_1^2
-2 y_\tau^2 ( p- n_{H_d} - n_L - n_E ) \delta_{i3}, \nonumber \\
\theta_{E,i} &=& \frac{12}{5} g_1^2
- 4 y_\tau^2 (p-n_{H_d} -n_L -n_E ) \delta_{i3}, \nonumber \\
\theta_{H_u} &=& 3 g_2^2 + \frac{3}{5} g_1^2
- 6 y_t^2 ( p- n_{H_u} - n_Q - n_U ), \nonumber \\
\theta_{H_d} &=& 3 g_2^2 + \frac{3}{5} g_1^2
-6 y_b^2 ( p- n_{H_d} - n_Q - n_D )
-2 y_\tau^2 ( p- n_{H_d} - n_L - n_E ).
\end{eqnarray}
As in the case of $\gamma_i$, $\theta^\prime_i$ is the same as $\theta_i$.

\end{document}